\documentstyle[amsmath,amssymb,graphicx]{article}

\def\nn{ }

\def\p{\partial}
\def\tr{{\rm tr}\,}
\def\Tr{{\rm Tr}\,}

\def\fund{{_\Box}}
\def\l[{\phantom.[}

\def\={=}
\def\||{||}

\def\={\
\begin{picture}(12,10)(0,0)
\put(0,0){\line(1,0){10}}
\put(0,5){\line(1,0){10}}
\end{picture}
\ }
\def\||{
\begin{picture}(10,10)(-3,3)
\put(0,1){\line(0,1){10}}
\put(7,1){\line(0,1){10}}
\end{picture}
\ }

\def\={\
\begin{picture}(12,10)(0,3)
\qbezier(0,0)(5,5)(10,0)
\qbezier(0,10)(5,5)(10,10)
\end{picture}
\ }
\def\||{
\begin{picture}(10,10)(-3,3)
\qbezier(0,0)(5,5)(0,10)
\qbezier(10,0)(5,5)(10,10)
\end{picture}
\ \ }

\def\ho{\overline}
\makeatletter
\newdimen\normalarrayskip              
\newdimen\minarrayskip                 
\normalarrayskip\baselineskip
\minarrayskip\jot
\newif\ifold             \oldtrue            \def\new{\oldfalse}
\def\arraymode{\ifold\relax\else\displaystyle\fi} 
\def\eqnumphantom{\phantom{(\theequation)}}     
\def\@arrayskip{\ifold\baselineskip\z@\lineskip\z@
     \else
     \baselineskip\minarrayskip\lineskip2\minarrayskip\fi}
\def\@arrayclassz{\ifcase \@lastchclass \@acolampacol \or
\@ampacol \or \or \or \@addamp \or
   \@acolampacol \or \@firstampfalse \@acol \fi
\edef\@preamble{\@preamble
  \ifcase \@chnum
     \hfil$\relax\arraymode\@sharp$\hfil
     \or $\relax\arraymode\@sharp$\hfil
     \or \hfil$\relax\arraymode\@sharp$\fi}}
\def\@array[#1]#2{\setbox\@arstrutbox=\hbox{\vrule
     height\arraystretch \ht\strutbox
     depth\arraystretch \dp\strutbox
     width\z@}\@mkpream{#2}\edef\@preamble{\halign
\noexpand\@halignto
\bgroup \tabskip\z@ \@arstrut \@preamble \tabskip\z@ \cr}%
\let\@startpbox\@@startpbox \let\@endpbox\@@endpbox
  \if #1t\vtop \else \if#1b\vbox \else \vcenter \fi\fi
  \bgroup \let\par\relax
  \let\@sharp##\let\protect\relax
  \@arrayskip\@preamble}
\def\eqnarray{\stepcounter{equation}%
              \let\@currentlabel=\theequation
              \global\@eqnswtrue
              \global\@eqcnt\z@
              \tabskip\@centering
              \let\\=\@eqncr

 \halign to \displaywidth\bgroup
    \eqnumphantom\@eqnsel\hskip\@centering
    $\displaystyle \tabskip\z@ {##}$%
    \global\@eqcnt\@ne \hskip 2\arraycolsep
         $\displaystyle\arraymode{##}$\hfil
    \global\@eqcnt\tw@ \hskip 2\arraycolsep
         $\displaystyle\tabskip\z@{##}$\hfil
         \tabskip\@centering
    &{##}\tabskip\z@\cr}
\newfont{\hr}{msbm10}
\newfont{\ams}{msam10}

\def\beq{\begin{equation}}
\def\eeq{\end{equation}}
\def\ba{\beq\new\begin{array}{c}}
\def\ea{\end{array}\eeq}
\def\be{\ba}
\def\ee{\ea}

\newdimen\linethick  \linethick=0.4pt
\newdimen\hboxitspace    \hboxitspace=5pt
\newdimen\vboxitspace    \vboxitspace=5pt

\def\fr#1{%
\beq\new
\vcenter{
\hrule height\linethick
          \hbox{\vrule width\linethick
                \kern\hboxitspace
                \vbox{\kern\vboxitspace
                      \hbox{$\begin{array}{c}\displaystyle#1
         \end{array}$}%
                      \kern\vboxitspace}%
                \kern\hboxitspace
                \vrule width\linethick}%
          \hrule height\linethick}%
\eeq}

\hoffset= - 1.0in         

\title{{\bf Evolution method and "differential
hierarchy" of colored knot polynomials } \vspace{.2cm}}
\author{{\bf A.Mironov}\footnote{ {\small {\it
Lebedev Physics Institute} and {\it ITEP, Moscow, Russia}};
mironov@itep.ru; mironov@lpi.ru}, {\bf A.Morozov}\thanks{{\small
{\it ITEP, Moscow, Russia}}; morozov@itep.ru}, {\bf
And.Morozov}\thanks{{\small {\it Moscow State University} and
{\it ITEP Moscow, Russia}};
Andrey.Morozov@itep.ru}\date{ }}

\begin{document}
 \maketitle

\vspace{-6.5cm}

\begin{center}
\hfill FIAN/TD-08/13\\
\hfill ITEP/TH-17/13\\
\end{center}

\vspace{5cm}

\centerline{ABSTRACT}

\bigskip

{\footnotesize
We consider braids with repeating patterns inside arbitrary knots
which provides a multi-parametric family of knots,
depending on the "evolution" parameter, which controls
the number of repetitions.
The dependence of knot (super)polynomials on such evolution parameters
is very easy to find.
We apply this evolution method to study of the families of knots
and links which include the cases with just two parallel and
anti-parallel strands in the braid,
like the ordinary twist and 2-strand torus knots/links
and counter-oriented 2-strand links.
When the answers were available
before, they are immediately reproduced,
and an essentially new example is added of the "double braid",
which is a combination
of parallel and anti-parallel 2-strand braids.
This study helps us to reveal with the full clarity
and partly investigate a mysterious
hierarchical structure of the colored HOMFLY polynomials,
at least, in (anti)symmetric representations, which extends the original
observation for the figure-eight knot to many (presumably all) knots. We demonstrate that this structure is typically respected by
the $t$-deformation to the superpolynomials.
}

\section{Introduction}

Knot polynomials \cite{knotpol} are Wilson loop averages in $3d$ Chern-Simons
theory \cite{CS,WitJones} sometimes further
deformed by refinement ($t$-deformation) procedures. They are non-trivial
generalizations of group characters and associated generating
functions generalize both the KP/Toda $\tau$-functions and the $2d$
conformal blocks. They are directly related to representation theory of
quantum algebras and, probably, after the $t$-deformation, of
double affine Hecke algebras. All this makes the study of knot
polynomials very interesting and important. The most important class
of HOMFLY polynomials has a straightforward definition \cite{TR}
in terms of traces of products of quantum ${\cal R}$-matrices in
knot diagrams treated as closed braids, which are further reduced to traces in the space of
intertwining operators, which no longer contain any reference to
a particular Lie algebra $SU_q(N)$, see \cite{MMMkn1,MMMkn2} and
references therein. Somewhat surprisingly, despite simplicity of
the definition, the evaluation of particular knot polynomials and the
study of extremely rich set of interrelations between them turns to
be a very sophisticated problem. A tremendous progress has been achieved
during the last years with the help of increased computer capacities,
but they are still far from being sufficient to study really
interesting questions. Only a combination of difficult computer
experiments with more traditional methods of theoretical physics
provides a key to this kind of problems. In this paper we elaborate
on one of such methods (evolution method) and discuss one of its
immediate implications (differential hierarchy), which, as often
happens, goes far beyond the applicability range of the method
itself.

The story begins with \cite{AS,DMMSS}, where it was suggested  to study
the knot polynomials for the one-parametric families of knots, since the
"evolution" along the discrete parameter can be often described in a
relatively simple way. This approach  was successfully used in
\cite{DMMSS} to construct the superpolynomials of the torus knots $T^{[m,n]}$,
where $n$ plays the role of the evolution parameter with $m$ fixed:
this method allows one to generalize the archetypical Rosso-Jones
formula \cite{RJ} to the superpolynomials (see also \cite{HLsuper}).
Moreover, in application to the torus knots the evolution method was
promoted by I.Cherendnik \cite{Che} (see also \cite{net}) to an
exhaustive solution to the problem of constructing the torus superpolynomial.
 Later the same
method was applied \cite{IMMM} in an "orthogonal" direction: to study
the figure-eight knot superpolynomials in various symmetric
representations $[r]$, in this case the evolution parameter is $r$.
Of course, this method works in just the same way for all other
twist knots $K_n$ (the figure-eights is $K_{-1}$), this
generalization was made in \cite{indtwist,FGSS}.
The two directions can be unified: one can consider the
two commuting evolutions, in two directions $n$ and $r$, both for
the families of torus and twist knots. The complexity of the problem
for the twist series is the same as that of the $2$-strand knots. This
is one of the example in this paper: formally, it adds nothing
essentially new to \cite{indtwist}, but we put the accents somewhat
differently.

More important, there is no need to stop at the level of ordinary
twist knots. Equally easy one can study more general families of what
we sometimes call generalized twist knots: we unify under this name all effectively
2-strand series, see below. In particular, funny double braid series are
provided by two "orthogonal" 2-braids.

The study of particular evolutions can look just as a technical
exercise, but in this way one can study relations between different
knot polynomials, with different knots and representations
\cite{MMeqs}. One day the sophisticated system of commuting
evolutions would lead to an understanding of a new interesting
integrable structure behind the knot/Chern-Simons theory in $3d$:
much more rich and interesting than the KP/Toda integrability,
governing the holomorphic $2d$ (thus, in some sense the $1d$) field
theories.

\paragraph{Notations.} Throughout the paper we use the notations:
\be \{x\}\equiv x-{1\over x}\ \ \ \ \ \ \ \ \ \ [x]\equiv
[x]_q\equiv {q^x-q^{-x}\over q-q^{-1}}\ \ \ \ \ \ \ \ \ \
(x)_k\equiv\prod_{i=0}^{k-1}\{xq^{i}\}\ \ \ \ \ \ \ \ \ \
(x)_k^*\equiv\prod_{i=0}^{k-1}\{xq^{-i}\} \ee We also associate with
the irreducible representation of $SU_q(N)$ (which we only consider
in the paper) the corresponding Young diagram $Q$ with the line
lengths $q_1\ge q_2\ge\ldots$ and $\nu_Q\equiv\sum_i(i-1)q_i$,
$\kappa_Q=\sum_{i,j\in Q}(j-i)= \nu_{Q'}-\nu_Q$, where $Q'$ denotes
the transposed Young diagram. We also denote $S_Q(p)$ the Schur
function of the Young diagram $Q$, which is equal to the character
of the representation $Q$, and $D_Q$ is its quantum dimension. $S_Q^*$ is
the value of $S_Q(p)$ at "the topological locus" $p_k = p_k^*
= \frac{\{A^k\}}{\{q^k\}}$.
At last, we use the ordinary letters for
the knot polynomials "on-shell" and the calligraphic letters for the
"off-shell" knot polynomials.

\section{Evolution method}

\subsection{Idea of the evolution method: 2-strand torus knots and links\label{21}}

\subsubsection{Fundamental representation}

Let us begin from the basic example of the fundamental HOMFLY
polynomial for the 2-strand knots and links $[2,n]$ (for odd $n$ we
get a knot, while for even $n$ a two-component link).

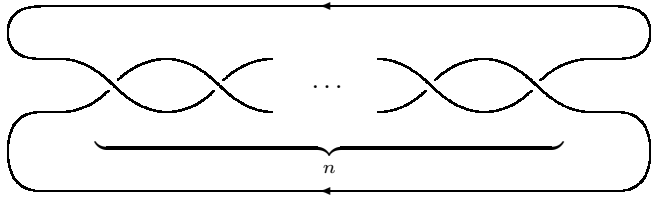
\begin{picture}(200,100)(180,-50)\label{2str}
\put(253,10){\line(1,0){7}} \put(253,-10){\line(1,0){7}}
\qbezier(260,10)(270,10)(280,0) \qbezier(280,0)(290,-10)(300,-10)
\qbezier(300,10)(310,10)(320,0) \qbezier(320,0)(330,-10)(340,-10)
\qbezier(260,-10)(270,-10)(278,-2) \qbezier(281.5,2)(290,10)(300,10)
\qbezier(300,-10)(310,-10)(318,-2) \qbezier(321.5,2)(330,10)(340,10)
\put(355,-1){\mbox{$\ldots$}}
\put(460,10){\line(1,0){10}} \put(460,-10){\line(1,0){10}}
\qbezier(380,10)(390,10)(400,0) \qbezier(400,0)(410,-10)(420,-10)
\qbezier(420,10)(430,10)(440,0) \qbezier(440,0)(450,-10)(460,-10)
\qbezier(380,-10)(390,-10)(398,-2) \qbezier(401.5,2)(410,10)(420,10)
\qbezier(420,-10)(430,-10)(438,-2) \qbezier(441.5,2)(450,10)(460,10)
\put(273,-20){\mbox{$\underbrace{\phantom{\longrightarrow\longrightarrow
\longrightarrow\longrightarrow\longrightarrow\longrightarrow\longrightarrow
\longrightarrow\longrightarrow\longrightarrow\longrightarrow }}_n$}}
\qbezier(253,10)(240,10)(240,20) \qbezier(253,30)(240,30)(240,20)
\qbezier(470,10)(483,10)(483,20) \qbezier(470,30)(483,30)(483,20)
\put(470,30){\line(-1,0){217}} \put(360,30){\vector(-1,0){2}}
\qbezier(253,-10)(240,-10)(240,-25)
\qbezier(253,-40)(240,-40)(240,-25)
\qbezier(470,-10)(483,-10)(483,-25)
\qbezier(470,-40)(483,-40)(483,-25) \put(470,-40){\line(-1,0){217}}
\put(360,-40){\vector(-1,0){2}}
\end{picture}

As a function of $n$ the HOMFLY polynomial for 2-strand knot in the
fundamental representation is a linear combination of just two
monomials (see ss.\ref{torus}):
\be\label{2}
H_{_\Box}^{[2,n]}(A,q) =\alpha_{_{\Box,0}}\!\left(-\frac{1}{Aq}\right)^n+
\alpha_{_{\Box,1}}\!\left(\frac{q}{A}\right)^n
\ee
where
$\alpha_{_{\Box,i}}$ do not depend on $n$ (but can
be functions of $A$ and $q$). These two coefficients can be easily
found, if one looks at two particular values $n=\pm 1$: in both cases
one gets the unknot, what means that
\be
\frac{q\alpha_{_{\Box,1}}}{A} -
\frac{\alpha_{_{\Box,0}}}{qA} = \frac{A\alpha_{_{\Box,1}}}{q} -
qA\alpha_{_{\Box,0}} = H_{_\Box}^{unknot}
\ee
and
\be
\alpha_{_{\Box,0}}
=\frac{\{A/q\}}{\{q^2\}}\,H_{_\Box}^{unknot} \ \ \ \ \ \ \ \ \ \ \ \
\ \ \ \ \ \ \alpha_{_{\Box,1}}
=\frac{\{Aq\}}{\{q^2\}}\,H_{_\Box}^{unknot}
\ee
Here and below we
denote by $H$ the reduced polynomials, so that $\
H_{_\Box}^{unknot}=1$, and
\be
H_{_\Box}^{[2,n]}(A,q) =
\frac{1}{\{q^2\}A^n} \Big(\{Aq\}q^n +(-)^n\{A\!/q\}q^{-n}\Big)
\label{H1tor2}
\ee

As a simple check of (\ref{H1tor2}), one can look at the case of
$n=0$, when there are two unlinked unknots so that the {\it unreduced} HOMFLY polynomial
is a product of two
\be
\check H_R^{[2,0]} = \Big(\check
H_R^{unknot}\Big)^2
\ee
and observe that the reduced one is
\be
H_R^{[2,0]} =
\frac{\Big(\check H_R^{unknot}\Big)^2}{\check H_R^{unknot}} = \check
H_R^{unknot} = D_R = S_R^*
\ee
In the fundamental representation
$R=[1]=\Box$, the quantum dimension $D_R$, i.e. the value $S_R^*$ of
the Schur polynomial $S_R\{p\}$ at "the topological locus" $p_k = p_k^*
= \frac{\{A^k\}}{\{q^k\}}$ is just
\be
\check H_{_\Box}^{unknot} =
D_{_\Box} = S_{_\Box}^*=p_1^*=\frac{\{A\}}{\{q\}}
\ee
and therefore
one expects that $H_R^{[2,0]} = \frac{\{A\}}{\{q\}}$. This is, indeed,
true for (\ref{H1tor2}):
\be
H_R^{[2,0]} =
\frac{\{Aq\}+\{A/q\}}{\{q^2\}} = \frac{(q+q^{-1})\{A\}}{q^2-q^{-2}}
= \frac{\{A\}}{\{q\}}
\ee

\subsubsection{A first summary}

This example contains all the essence of the evolution method:

\begin{itemize}
\item pick up a parameter $n$, on which the answer depends in
a controllable way,
\item define the remaining $n$-independent parameters from "initial
conditions", i.e. from the known answers at particular values of $n$,
\item if answers are known in more cases than there are parameters in
the ansatz, use them to check the resulting formula.
\end{itemize}

Of course, all this depends on whether one knows anything about the
$n$-dependence and possesses enough many known formulas to be used as
initial conditions. Both these conditions depend on the clever
choice of the evolution parameter $n$.

Before we discuss when and why this can be possible, we elaborate a
little further on immediate generalizations of (\ref{H1tor2}) in
three directions: higher representations, more strands and
$t$-deformation from the HOMFLY polynomials to the superpolynomials.

\subsubsection{Symmetric representation $[2]$}

In this case the $n$-dependence is provided by the $3$-term formula:
\be
H_{[2]}^{[2,n]} =   \frac{1}{A^nq^{4n}}\Big( \alpha_{[2],0}+
\alpha_{[2],1}(-q^2)^n +\alpha_{[2],2}q^{6n}\Big) \label{H2tor2}
\ee
and the initial conditions are provided by the three known cases of $n=-1,0,1$:
\be
H_{[2]}^{[2,\pm 1]} = 1, \nn \\
H_{[2]}^{[2,0]} = D_{[2]} = S_{[2]}^* =
\frac{\{Aq\}\{A\}}{\{q^2\}\{q\}}
\ee
so that
\be\label{12}
H_{[2]}^{[2,n]} =
\frac{1}{A^nq^{4n}}\left(\frac{\{A\}\{A/q\}}{\{q^4\}\{q^2\}}
 +
\frac{\{Aq^2\}\{A/q\}}{\{q^4\}\{q\}}(-q^2)^n +
\frac{\{Aq^3\}\{Aq^2\}}{\{q^4\}\{q^3\}}q^{6n}\right)
\ee

Similarly, in the case of the antisymmetric representation $[11]$,
the ansatz is
\be
H_{[11]}^{[2,n]} = \frac{1}{A^nq^{4n}}\Big(
\alpha_{[11],0}+
\alpha_{[11],1}(-q^2)^n +\alpha_{[11],2}q^{6n}\Big)
\ee
the initial conditions are
\be
H_{[11]}^{[2,\pm 1]} = 1, \nn \\
H_{[11]}^{[2,0]} = D_{[11]} = S_{[11]}^* =
\frac{\{A/q\}\{A\}}{\{q^2\}\{q\}}
\ee
and the final answer is
\be
H_{[11]}^{[2,n]}(A,q) = \frac{1}{A^nq^{4n}}\left(
\frac{\{A\}\{A/q\}}{\{q^4\}\{q^2\}} +
\frac{\{A/q^2\}\{Aq\}}{\{q^4\}\{q\}}(-q^2)^{-n} +
\frac{\{A/q^3\}\{A/q^2\}}{\{q^4\}\{q^3\}}q^{-6n}\right) =
H_{[2]}^{[2,n]}(A,q^{-1})
\ee

\subsubsection{Superpolynomial in the fundamental representation}

The relevant ansatz in this case is different for knots and for
links. In the case of odd $n$ (knots) it is:
\be
P_{_\Box}^{[2,n]} =
\frac{1}{A^n}\Big(\alpha_{_\Box}t^n - \beta_{_\Box}q^{-n}\Big), \ \
\ \ \ n\ {\rm odd} \label{P1tor2}
\ee
and the initial conditions at
$n=\pm 1$ are
\be
P_{_\Box}^{[2,\pm 1]} = P_{_\Box}^{unknot} = 1
\ee
This implies that
\be
\alpha_{_\Box} t - \beta_{_\Box}/q = A, \nn \\
\alpha_{_\Box}/ t - \beta_{_\Box}q = 1/A \ee and \be
P_{_\Box}^{[2,n]} = \frac{1}{A^n\{qt\}}\Big(\{Aq\}t^n -
\{A/t\}q^{-n}\Big)
\ee

\subsubsection{3-strand torus knots and links}

This time the parameter $n$ counts the number of "torus wrappings" and
\be
H_{_\Box}^{[3,n]} = \frac{1}{A^{2n}}\Big(\alpha_{_\Box}^{(3)}
q^{2n} +\beta_{_\Box}^{(3)} (e^{2\pi in/3} + e^{-2\pi in/3})
+\gamma_{_\Box}^{(3)} q^{-2n}\Big) \label{H1tor3})
\ee
The three
initial conditions at $n=\pm 1$ (unknots) and $n=0$ (three
independent unknots) imply:
\be
\alpha q^2 - \beta + \gamma q^{-2} = A^2, \nn \\
\alpha q^{-2} - \beta + \gamma q^{2} = A^{-2}, \nn \\
\alpha  +2 \beta + \gamma   = \left(\frac{\{A\}}{\{q\}}\right)^2
\ee
Solving these equations and substituting them back into the ansatz,
one obtains
\be
H_{_\Box}^{[3,n]} =
\frac{1}{A^{2n}\{q^3\}\{q^2\}}\Big(\{Aq^2\}\{Aq\}\, q^{2n} +
2(q+1/q) \cos\frac{2\pi n}{3}\{Aq\}\{A/q\} +\{A/q^2\}\{A/q\}\,
q^{-2n}\Big)
\ee

\subsubsection{A second summary and comments}

This set of examples illustrates the fact that the evolution method can
be used in a large variety of cases. However, before we proceed and
further extend this variety, it is necessary to explain the origin
of the above ansatze. In this presentation we follow the papers
\cite{TR,DMMSS} and especially \cite{MMMkn1,MMMkn2}.

In fact, the HOMFLY polynomial for a 2-strand knot/link like
(\ref{H1tor2}), is the weighted trace of the $n$-th power of the
quantum ${\cal R}$-matrix in representation $R$:
\be
H_R^{[2,n]} =
{\rm Tr}_{R\times R} {\cal R}^n
\ee
The representation $\ R\times R =
\oplus\, Q\ $ is reducible and can be decomposed into a combination of
irreducible representations $Q$. In each of these representations
the ${\cal R}$-matrix acts as unit matrix with a $Q$-dependent
eigenvalue $\lambda_Q$, so that
\be
H_R^{[2,n]} = \sum_{Q\in
R\otimes R} \lambda_Q^n D_Q \label{HRtor2}
\ee
where $D_Q = {\rm
Tr}_Q I$ is a weighted trace of unity in representation $Q$ (for
$q=1$ and for the group $SU(N)$ this is just the ordinary
dimension of representation $Q$, but we rather express its quantum
deformation at $q\neq 1$ through the universal HOMFLY variable
$A=q^N$). The eigenvalues $\lambda_Q$ are given by the simple
formula,
\be
\lambda_Q = f_R\cdot\epsilon_Q q^{\varkappa_Q}
\label{lamQ}
\ee
where $\ \epsilon_Q = \pm 1\ $ and $\ \varkappa_Q =
\sum_{(i,j)\in Q}(j-i)\ $ is the eigenvalue of the cut-and-join
operator \cite{CJ,MMN}
\be
\hat W[2] = \frac{1}{2}\sum_{a,b}
\left((a+b)p_ap\,_b\frac{\p}{\p p_{a+b}} + abp_{a+b}\frac{\p^2}{\p
p_a\p p\,_b}\right)
\ee
when this latter acts on its eigenfunction, the
Schur function $S_Q\{p\,\}$. In (\ref{lamQ}) we introduced also
a normalization factor $f_R = q^{-4\varkappa_R}A^{-|R|}$, which
depends not on $Q$, but on the original $R$: it is important for
making the answer for the HOMFLY polynomial topological invariant.

Eq.(\ref{HRtor2}) is behind our first two ansatze (\ref{H1tor2}) and
(\ref{H2tor2}),  with $\ \varkappa_{_\Box}\!=0;\ \
\varkappa_{[2]}=-\varkappa_{[11]}=1;\ \ \varkappa_{[4]}=6,\ \
\varkappa_{[31]}=2,\ \ \varkappa_{[22]}=0\ $ and alternating signs
$\epsilon_Q$ within every family $R\otimes R$. The slight difference
is that we restored the coefficients $D_Q$ by the evolution method,
while in (\ref{HRtor2}) they are explicitly known. We did so for
illustrative purposes: of course, the 2-strand knots are not a big
mystery, but they can be used as a sample example to illustrate
the method, in cases where an explicit formula like (\ref{HRtor2}) is
not yet available.

An example of this kind is (\ref{P1tor2}): it is a guess of
\cite{DMMSS} that the eigenvalues $q^{\varkappa_Q}$ are substituted
by $t^{\nu_{\bar Q}}q^{-\nu_{Q}}$ where $\nu_Q$ are the well-known
constituents of $\varkappa_Q = \nu_{\bar Q} - \nu_Q$ and $\nu_Q =
\sum_i (i-1)q_i$ for $Q = \{q_1\geq q_2\geq \ldots \geq 0\}$, though
the corresponding counterpart of the set of the cut-and-join operators
\cite{MMN} is yet unknown, nothing to say about the
$t$-deformation of the ${\cal R}$-matrix structure behind
(\ref{HRtor2}). In absence of this knowledge, there is no way to
{\it deduce} the coefficients like $D_Q$, and the evolution method at
the moment remains the most powerful approach to the torus
superpolynomials (for the torus HOMFLY polynomial there is a general Rosso-Jones
formula \cite{RJ,chi,BEM}).

In the HOMFLY case, the $n$-evolution of the torus knots $[m,n]$
depends not only on representation $R$, as illustrated by the
difference between (\ref{H1tor2}) and (\ref{H2tor2}), but also on
the number of strands $m$, see (\ref{H1tor3}). Still, for any given
$m$ there is no difference between, say, knots and links: the
formula is the same, independently of whether $m$ and $n$ are
coprime or not.
Remarkably, as found in \cite{DMMSS}, the $t$-deformation further splits
this universal evolution into branches, which depend on the residue
$\ n\ {\rm mod}(m)$: already for the $2$-strand case the evolution
of knots and link superpolynomials is governed by different
coefficients $\alpha_{_{\Box,i}}$. In practice, this
means that one needs even more initial conditions to fix the larger
set of parameters.

The last comment is once again about eq.(\ref{H1tor3}). This time
the evolution is associated with adding of two crossings (two ${\cal
R}$-matrices) instead of one. In other words, it is controlled not
by the eigenvalue of ${\cal R}$, but by that of the product ${\cal
R}_1{\cal R}_2$, where ${\cal R}_s$ stands at the crossing of
strands with numbers $s$ and $s+1$ in the braid. The matrices ${\cal
R}_s$ and ${\cal R}_{s+1}$ do not commute, so the eigenvalue of the
product is {\it not} a product of the two eigenvalues and has to be
calculated, see \cite{MMMkn1,MMMkn2}. Particular products
$\prod_{s=1}^{m-1}{\cal R}_s$, relevant for the torus knots have very
simple eigenvalues, $\epsilon_Q q^{2\varkappa_Q/m}$ (these are the
ones, appearing in the Rosso-Jones formula), but in more general
situations the eigenvalues can be far more complicated (see
\cite{MMMkn2}, for some examples).

\subsection{General formulation of evolution method: off-shell evolution}

Now we are ready to formulate more accurately what is the evolution
method. Let us cut a knot diagram ${\cal D}$, representing some
knot/link into two pieces, $E$ and $B$, so that in both cases the
equal number $m$ of lines is cut. Then consider a family ${\cal
D}_n$, where $B$ is substituted by a chain of $n$ copies of $B$.
Conceptually, the knot polynomials for this chain are
\be
{\cal H}_R^{{\cal
D}_n} = \Tr_{R^{\otimes m}} {\cal E}{\cal B}^n \label{Hevo}
\ee
where ${\cal E}$ and ${\cal B}$ are the products
of ${\cal R}$-matrices inside the blobs $A$ and $B$, and the trace also includes the factor $q^\rho$
\cite{TR,MMMkn1,MMMkn2}

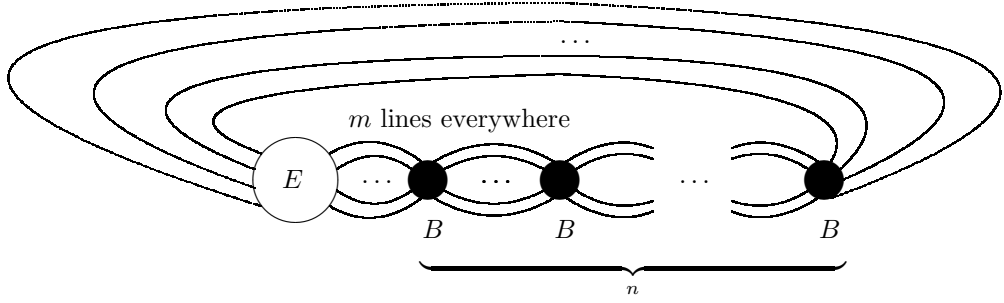
\begin{picture}(300,150)(-200,-50)
\put(-100,0){\circle{30}} \put(-50,0){\circle*{20}}
\put(0,0){\circle*{20}} \put(45,-1){\mbox{$\ldots$}}
\put(100,0){\circle*{20}}
\put(-53,-30){\mbox{$\underbrace{\phantom{\longrightarrow\longrightarrow
\longrightarrow\longrightarrow\longrightarrow\longrightarrow\longrightarrow
\longrightarrow\longrightarrow\longrightarrow }}_n$}}
\put(-80,20){\mbox{$m\ {\rm lines\ everywhere}$}}
\qbezier(-87,10)(-70,20)(-52.5,7) \qbezier(-84.5,4)(-70,14)(-56,4)
\put(-75,-1){\mbox{$\ldots$}} \qbezier(-87,-10)(-70,-20)(-52.5,-7)
\qbezier(-84.5,-4)(-70,-14)(-56,-4)
\qbezier(-47.5,7)(-25,20)(-2.5,7) \qbezier(-44,4)(-25,14)(-6,4)
\put(-30,-1){\mbox{$\ldots$}} \qbezier(-47.5,-7)(-25,-20)(-2.5,-7)
\qbezier(-44,-4)(-25,-14)(-6,-4)
\qbezier(2.5,7)(20,18)(35,12) \qbezier(6,4)(20,13)(35,8)
\put(-30,-1){\mbox{$\ldots$}} \qbezier(2.5,-7)(20,-18)(35,-12)
\qbezier(6,-4)(20,-13)(35,-8)
\qbezier(97.5,7)(80,18)(65,12) \qbezier(94,4)(80,13)(65,8)
\put(-30,-1){\mbox{$\ldots$}} \qbezier(97.5,-7)(80,-18)(65,-12)
\qbezier(94,-4)(80,-13)(65,-8)
\qbezier(-113,10)(-180,35)(-0,40) \qbezier(102.5,7)(120,35)(-0,40)
\qbezier(-115.5,4)(-220,45)(-0,47) \qbezier(106,4)(150,45)(-0,47)
\put(0,52){\mbox{$\ldots$}} \qbezier(-113,-10)(-350,65)(-0,67)
\qbezier(102.5,-7)(270,55)(-0,67)
\qbezier(-115.5,-3)(-280,55)(-0,60) \qbezier(107,0)(220,50)(-0,60)
\put(-105,-3){\mbox{$E$}} \put(-52,-22){\mbox{$B$}}
\put(-2,-22){\mbox{$B$}} \put(98,-22){\mbox{$B$}}
\end{picture}

Now we can attach to this diagram an $m$-strand braid (one can
attach something else, e.g. several braids with $m_1+\ldots = m$,
but this adds nothing new to our consideration), and there are
different options for orientation: in result, there will be $m_L$
strands in representation $R$ and $m_R$ strands in the conjugate
representation $\bar R$, $m_R+m_L = m$. In
fact, according to the general strategy of \cite{MMMkn1,MMMkn2}, the giant
sum (trace) over the representation space
\be\label{tp}
R^{\otimes m_L}\otimes \bar
R^{\otimes m_R} = \oplus_Q {\cal M}_{R\bar R}^Q \otimes Q
\ee
in
(\ref{Hevo}) reduces to the one over a relatively smaller set of
the intertwining operators ${\cal M}_{R\bar R}^Q $. Let us see how it works.

The both factors ${\cal E}$ and ${\cal B}$ can be expanded w.r.t. to the second factor in (\ref{tp})
into irreducible representations
$Q\vdash m|R|$, with eigenvalues $E_Q$ and $B_Q$ respectively,
and the answer for {\it extended} or {\it off-shell} HOMFLY polynomial is
\be\label{ans}
{\cal H}^{{\cal E}{\cal B}^n}_R\{p\} = \sum_{Q\vdash m|R|} C_{RQ}\Big(\Tr_{{\cal M}_{R\bar R}^Q} E_QB_Q^n\Big) S_Q\{p\}
\ee
where $\Tr_{{\cal M}_{R\bar R}^Q}$ means the trace over the space of intertwining operators (which is generally not
one-dimensional if $m>2$).
Here $C_{RQ}$ are the multiplicities in the expansion:
\be
C_{RQ} = {\rm dim} {\cal M}_{R\bar R}^Q
\ee
The result can be easily put on-shell by choosing $p_k$ in (\ref{ans}) to lie on the topological locus.

In particular, for the torus evolution ${\cal E}=I$ one gets the Rosso-Jones formula \cite{RJ}
\be
{\cal H}^{T[m,n]}_R = \sum_{Q\vdash m|R|} \epsilon_Q c_{RQ} q^{2\varkappa_Q n/m} S_Q\{p\}
\ee
with the expansion coefficients $c_{RQ}$ determined by the Adams operation $\widehat{Ad}_m F\{p_k\} = F\{p_{mk}\}$:
\be\label{RJ1}
\widehat{Ad}_m S_R\{p\} = \sum_{Q\vdash m|R|} c_{RQ} S_Q\{p\}
\ee

Once again, for this to work one needs to know/guess the $B_Q$ and
sufficient number of initial conditions. However, even before that
the r.h.s. of (\ref{ans}) is a non-trivial assumption. The point is
that the middle part of the formula is not yet known in any
constructive form not only for the superpolynomials (where ${\cal E}$ and ${\cal B}$
should be elements of something like a double affine Hecke algebra
(DAHA) \cite{Che} rather than ordinary matrices), even for the HOMFLY polynomials
there are reservations. The point is that an absolutely reliable
formalism exists \cite{TR} only for braids, not for arbitrary knot
diagrams (also for the Jones polynomials, where the Kauffman matrix
\cite{Kaufm}, which respects all the three Reidemeister moves is
explicitly known, at least in the fundamental representation).
Therefore, it is not quite clear, what is the accurate expression for
$E_Q$ for such diagrams. The evolution method, and it is its big
advantage, simply ignores this gap, it derives $E_Q$ from just an
assumption that they exist, without using any kind of explicit
definition (nothing to say about a {\it constructive} one). This is
why we prefer to consider many results, obtained by the evolution
method as {\it conjectures}, but as time goes there are more and
more confirmations from various explicit calculations that the
results are correct. Moreover, the method provides them in numerous
examples and with far less effort than any other, and, perhaps,
they can indeed be trusted.

\subsection{Torus knots, extended knot polynomials and seesaw evolution}

\subsubsection{The iterative seesaw evolution of \cite{DMMSS}}

The archetypical example of evolution formula is provided by
the Rosso-Jones formula (\ref{RJ1}), \cite{RJ} for the torus HOMFLY polynomials,
which looks the best for the {\it extended} polynomials of \cite{DMMSS} and \cite{MMMkn1}:
\be
{\cal H}^{[m.n]}_R\{p\,\} = q^{\frac{2n}{m}\hat W} \widehat{\rm Ad}_m S_R\{p\,\}
\label{RJ21}
\ee
The Adams operation
looks even better after a Miwa transform $p_k = \tr X^k$, where
$X$ is some $M\times M$ matrix\footnote{
This transformation actually reduces the entire space of time variables
to a certain $M$-dimensional hyperspace.
The point is that, while all formulas in terms of $X$ are explicitly
dependent on $M$, they are being written for {\it functions}, like
$S_R\{p\}$ and ${\cal H}_R\{p\}$, which are $M$-independent.
This important technique is widely used since the theory of
Kontsevich matrix models  \cite{KoMamo}.
}:
then $p_{mk} = \tr X^{mk}$
and denoting $F\{p\} = F[X]$, one rewrites (\ref{RJ21}) as
\be
{\cal H}^{[m.n]}_R[X] = q^{\frac{2n}{m}\hat W} S_R[X^m]
\label{RJ2}
\ee
and the cut-and-join operator in this representation \cite{MMN} is simply
$\hat W = \frac{1}{2} \tr \left( X\frac{\p}{\p X^{tr}}\right)^2$.
Clearly, this formula looks like a {\it double} evolution in two
different directions: along $n/m$ and $m$.

Generalization of this formula to the superpolynomials is straightforward
\cite{AS,DMMSS,Che}, but very interesting.

First \cite{DMMSS}, the evolution operator $q^{\hat W}$ with the eigenfunctions $S_Q[X]$
and eigenvalues $q^{\varkappa_Q}$
is now substituted by $\hat U$ with eigenfunctions, which are the MacDonald
polynomials $M_Q[X]$, explicitly depending on two parameters $q$ and $t$,
and eigenvalues $q^{\nu_{\bar Q}}t^{-\nu_{Q}}$  (we return to the HOMFLY
case, when $t=q$ and $\varkappa_Q = \nu_{\bar Q} - \nu_Q$).

Second, while in the Rosso-Jones formula $\frac{n}{m}\varkappa_Q$ was
always integer (only Young diagrams $Q$ with this property were
actually contributing), this is no longer the case for the superpolynomials:
$\frac{n}{m}\nu_Q$ is not always integer.
However, as found in \cite{DMMSS} the evolution parameter is actually not the ratio
$n/m$: the evolution operator $\hat U$ is raised to another {\it integer} power $k$,
where $n = mk+r$.
Then, the $k$-evolution states that
\be
{\cal P}^{[m,n]}_R = \hat U^k {\cal P}^{[m,r]}_R
\ee
where the residues $r$ is now smaller than $m$.

Third, the ordinary knot polynomials should be the same
for knots $[m,r]$ and $[r,m]$, simply because these
are homotopically equivalent:
\be
P^{[m,r]}_R = P^{[r,m]}_R
\label{mrequiv}
\ee
Since $m>r$, one could once again use the evolution rule
and further reduce
\be
{\cal P}^{[r,m]}_R = \hat U^{k_1} {\cal P}^{[r,r_1]}_R
\ee
where now $m = kr+r_1$.
This procedure can be iteratively continued:
\be
\begin{array}{cccccccc}
{\cal P}^{[m,n]}_R &=  \hat U^k {\cal P}^{[m,r]}_R \\
&\downarrow \\
&  {\cal P}^{[r,m]}_R &=  \hat U^{k_1} {\cal P}^{[r,r_1]}_R \\
&&\downarrow \\
&&
{\cal P}^{[r_1,r]}_R &=  \hat U^{k_2} {\cal P}^{[r_1,r_2]}_R \\
&&&\downarrow \\
&&&& \ldots \\
&&&&&\downarrow \\
&&&&&{\cal P}^{[r_s,r_{s-1}}_R &=  \hat U^{k_{s+1}} {\cal P}^{[r_s,1]}_R \\
&&&&&&\downarrow \\
&&&&&&{\cal P}^{[1,r_s]}_R &= {\cal M}_R
\label{itescheme1}
\end{array}
\ee
where ${\cal M}_R$ is just the MacDonald polynomial, postulated
in \cite{DMMSS} to
describe the {\it unknot} superpolynomial.
The sequence of residues $n>m>r > r_1> r_2> \ldots > r_s>r_{s+1}=1$ is
provided by Euclid's algorithm for generating the maximal
common divisor $r_{s+1}=1$ of the two coprime integers $m$ and $n$:
\be
\begin{array}{cccccccccc}
n = &mk+r \\
& m = &rk_1 + r_1  \\
&& r = &r_1k_2+r_2 \\
&&& r_1 = & r_2k_3 + r_3 \\
&&&&\ldots \\
&&&&& r_{s-1} = &r_{s}k_{s+1} + 1  \\
\end{array}
\ee
This iterative evolution scheme  was successfully applied in \cite{DMMSS}
and further in \cite{HLhomfly,HLsuper}
to a variety of torus knots.
However, it was not fully satisfactory,
because the status of the horizontal equalities and the
vertical arrows in (\ref{itescheme1}) is different.
The horizontal equalities hold for the extended polynomials,
depending on time variables $\{p_k\}$ (or on the Miwa variables $X$),
but the vertical arrows are implied by (\ref{mrequiv}),
which is valid only on {\it the topological locus}
\be\label{tolo}
p_k = p_k^* = \tr X_*^k = \frac{\{A^k\}}{\{t^k\}}, \ \
{\rm i.e.\ for}  \ \ X_* = {\rm diag} \Big(t^{N(M+1-2j)} \ \Big| \ j=1,\ldots,M\Big)
\ee
Because of this {\it not} all of the coefficients in ${\cal P}^{[m,r]}_R\{p\}$
can be  unambiguously determined from (\ref{mrequiv}) for sufficiently big
knots and representations,
some additional procedure is needed to revert
the vertical arrows in (\ref{itescheme1}).

\subsubsection{ASC realization  of seesaw  evolution}

Such a procedure was successfully suggested in \cite{AS} and \cite{Che}
(it was partly inspired by the theory
of Verlinde algebras\footnote{
In particular the seesaw evolution rule (\ref{ssev}) below can be inspired
not only by (\ref{itescheme1}), but also by the $SL(2,{\bf Z})$ decomposition
$$
\left(\begin{array}{cc}  m & * \\ n & * \end{array}\right) =
\tau_{(-)^{s+1}}^{k}\tau_{(-)^{s}}^{k_{1}}
\ldots \tau_{+}^{k_{s-1}}\tau_{-}^{k_{s}}\tau_{+}^{k_{s+1}}\tau_{-}^{r_{s}}
$$
with $$\tau_- =
\left(\begin{array}{cc}  1 & 0 \\ 1 & 1 \end{array}\right), \ \ \ \ \ \ \ \ \
\tau_+ =
\left(\begin{array}{cc}  1 & 1 \\ 0 & 1 \end{array}\right)
$$
widely used in the old-fashioned theory of torus knots.
}).
Whatever the origin, the recipe is very simple.

{\bf First,} the recursion (\ref{itescheme1}) is reversed to provide the double evolution
\be
\boxed{
{\cal P}^{[m,n]}_R[X] =
\hat U_{[m,n]} {\cal M}_R =
\hat U_{(-)^s}^{k}\hat U_{(-)^{s+1}}^{k_{1}}
\ldots \hat U_{-}^{k_{s-1}}\hat U_{+}^{k_{s}}\hat U_{-}^{k_{s+1}} {\cal M}_R[X]
}
\label{ssev}
\ee
with two {\it different} evolution operators $\hat U_+$ and $\hat U_-$.
These {\it two} are what is inherited by the generic seesaw evolution from
the double one in (\ref{RJ2}).

For low values of $r$, considered in detail in \cite{DMMSS,HLhomfly,HLsuper},
the evolution operators are:
\be
\begin{array}{cccccccc}
r=1 & & \hat U_{[m,mk+1]} = \hat U_-^k \hat U_+^{m-1} \hat U_-, &&
r=2 &  m=2k_1 + 1 & \hat U_{[m,mk+2]} = \hat U_-^k \hat U_+^{k_1} \hat U_-^2 , \cr
r=3 & m=3k_1 \pm 1 &  \hat U_{[m,mk+3]} = \hat U_-^k \hat U_+^{k_1} \hat U_-^2, &&
\ {\rm or}    & m=3k_1 + 1 &  \hat U_{[m,mk+3]} = \hat U_-^k \hat U_+^{k_1} \hat U_-^3,  \cr
  &&&&  & m=3k_1 + 2 &  \hat U_{[m,mk+3]} = \hat U_-^k \hat U_+^{k_1} \hat U_-\hat U_+\hat U_-,\cr
r=4 & m=4k_1 \pm 1 &  \hat U_{[m,mk+4]} = \hat U_-^k \hat U_+^{k_1} \hat U_-^2,  &&
\ {\rm or}    & m=4k_1 + 1 &  \hat U_{[m,mk+4]} = \hat U_-^k \hat U_+^{k_1} \hat U_-^4,\cr  &&&&
    & m=4k_1 + 3 &  \hat U_{[m,mk+4]} = \hat U_-^k \hat U_+^{k_1} \hat U_-\hat U_+^2\hat U_-,\cr
r=5 & m=5k_1 \pm 1 &  \hat U_{[m,mk+5]} = \hat U_-^k \hat U_+^{k_1} \hat U_-^2,  &&
    & m=5k_1 \pm 3 &  \hat U_{[m,mk+5]} = \hat U_-^k \hat U_+^{k_1} \hat U_-^2,  \cr
r=6 & m=6k_1 \pm 1 &  \hat U_{[m,mk+6]} = \hat U_-^k \hat U_+^{k_1} \hat U_-^2,  &&
r=7 & m=7k_1 \pm 1 &  \hat U_{[m,mk+7]} = \hat U_-^k \hat U_+^{k_1} \hat U_-^2,  \cr
 & m=7k_1 \pm 2 &  \hat U_{[m,mk+7]} = \hat U_-^k \hat U_+^{k_1} \hat U_-^2,  &&
 & m=7k_1 \pm 3 &  \hat U_{[m,mk+7]} = \hat U_-^k \hat U_+^{k_1} \hat U_-^2,  \cr&\ldots&\end{array}
\ee

{\bf Second,} these operators possess different realizations,
the two well-known are
the Aganagic-Shakirov realization \cite{AS} in terms of
$S$ and $T$ generators of the $SL(2,{\bf Z})$ algebra in particular representations
and Cherednik's realization \cite{Che} in terms of the polynomial
representation of the double-affine Hecke algebra.
The latter approach led \cite{net} to a very explicit
representation of the $\gamma$-factors of \cite{DMMSS} through
sums over standard Young tableaux.

However, our goal is different: it is to explain what is the actual
way out of the uncertainty in the inversion of the vertical arrows
in (\ref{itescheme1}), which leads to the very possibility to {\it define}
the seesaw evolution (\ref{ssev}).

\subsubsection{Seesaw evolution in extended space of auxiliary links}

The basic idea \cite{Sham} is to extend consideration from the torus
knot superpolynomials to peculiar links, which involve a knot on the
surface of a torus in $S^3$ and an unknot inside or outside the
torus, going along the contractable $B$ or uncontractable $A$ cycles
respectively. We denote the corresponding link superpolynomials
through ${\cal P}^{[m,n]\otimes[0,1]}_{R\otimes Y}$ and ${\cal
P}^{[m,n]\otimes[1,0]}_{R\otimes Y}$ respectively. The idea is that
now one has an additional free parameter: the representation $Y$
associated with the auxiliary unknot, and in this extended space the
vertical arrows in (\ref{itescheme1}) will be reversible.

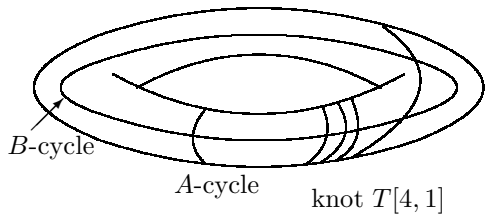
\begin{picture}(300,100)(-150,-50)
\qbezier(-60,-20)(0,-40)(60,-20)
\qbezier(-60,20)(-110,0)(-60,-20)
\qbezier(-60,20)(0,40)(60,20)
\qbezier(60,20)(110,0)(60,-20)
\qbezier(-55,5)(0,-25)(55,5)
\qbezier(-46,0)(0,25)(46,0)
%
\qbezier(-60,-10)(0,-30)(60,-10)
\qbezier(-60,10)(-90,0)(-60,-10)
\qbezier(-60,10)(0,30)(60,10)
\qbezier(60,10)(90,0)(60,-10)
\put(-95,-25){\mbox{$B$-cycle}}
\put(-86,-16){\vector(1,1){12}}
%
\qbezier(-20,-29)(-30,-20)(-20,-8)
\put(-32,-40){\mbox{$A$-cycle}}
%
%
\qbezier(18,-29)(30,-20)(24,-7)
\qbezier(24,-28)(38,-19)(30,-6)
\qbezier(30,-27)(42,-20)(35,-4)
\qbezier(36,-26)(80,0)(47,23)
\put(20,-45){\mbox{knot $T[4,1]$}}
\end{picture}

\noindent
It is assumed that the $B$-cycle lies inside, while
the $A$-cycle outside the torus, so that they are linked with the
knot, which is exactly on the torus surface.

The next picture shows the braid representation of the knot $T[4,1]$
and the $A$-cycle (to illustrate the structure we increase $n$ to $n=2$,
this literally means that, instead of the knot, there would be a 2-component link $T[4,2]$).
The $B$-cycle is also represented as an additional strand, if the circle is cut
and the two ends of the cut are pulled up and down.

Warning: In actual formulas the ordering and orientation of crossings with
the $A$ and $B$-cycles can be different from those shown in this picture.

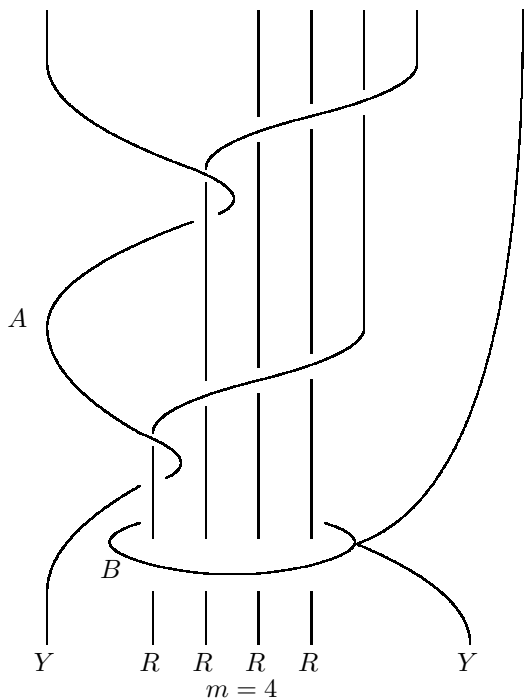
\begin{picture}(300,250)(-150,-60)
\put(0,-20){\line(0,1){50}}
\put(20,-20){\line(0,1){55}}
\put(40,-20){\line(0,1){60}}
\put(-20,-20){\line(0,1){35}}
\qbezier(-20,20)(-20,30)(20,40)
\qbezier(20,40)(60,50)(60,60)
\put(60,60){\line(0,1){20}}
\put(40,50){\line(0,1){30}}
\put(20,45){\line(0,1){35}}
\put(0,40){\line(0,1){40}}
\qbezier(-20,-30)(10,-37)(40,-30)
\qbezier(-25,-14)(-50,-22)(-20,-30)
\qbezier(45,-14)(70,-22)(40,-30)
\put(-40,-35){\mbox{$B$}}
\put(0,-40){\line(0,-1){20}}
\put(20,-40){\line(0,-1){20}}
\put(40,-40){\line(0,-1){20}}
\put(-20,-40){\line(0,-1){20}}
\put(-60,-60){\line(0,1){20}}
\qbezier(-60,-40)(-60,-20)(-25,0)
\qbezier(-15,3)(0,10)(-25,20)
\qbezier(-25,20)(-60,40)(-60,60)
\put(-75,60){\mbox{$A$}}
\put(20,80){\line(0,1){50}}
\put(40,80){\line(0,1){55}}
\put(60,80){\line(0,1){60}}
\put(0,80){\line(0,1){35}}
\qbezier(0,120)(0,130)(40,140)
\qbezier(40,140)(80,150)(80,160)
\put(80,160){\line(0,1){20}}
\put(60,150){\line(0,1){30}}
\put(40,145){\line(0,1){35}}
\put(20,140){\line(0,1){40}}
%
\qbezier(-60,60)(-60,80)(-5,100)
\qbezier(5,103)(20,110)(-5,120)
\qbezier(-5,120)(-60,140)(-60,160)
\put(-60,180){\line(0,-1){20}}
\put(-65,-70){\mbox{$Y$}}
\put(-25,-70){\mbox{$R$}}
\put(-5,-70){\mbox{$R$}}
\put(15,-70){\mbox{$R$}}
\put(35,-70){\mbox{$R$}}
\put(95,-70){\mbox{$Y$}}
\put(0,-80){\mbox{$m=4$}}
\qbezier(57,-22)(120,0)(120,180)
\qbezier(57,-22)(100,-40)(100,-60)
\end{picture}

\bigskip

\bigskip

\bigskip

\bigskip

The new scheme looks like
\be
\begin{array}{ccccccc}
{\cal P}^{[m,n]\otimes[0,1]}_{R\otimes Y} &=  \hat U^k {\cal P}^{[m,r]\otimes[0,1]}_{R\otimes Y} \\
&\downarrow \\
&  {\cal P}^{[r,m]\otimes[1,0]}_{R\otimes Y} &
=  \hat V^{k_1} {\cal P}^{[r,r_1]\otimes[1,0]}_{R\otimes Y} \\
&&\downarrow \\
&&
{\cal P}^{[r_1,r]\otimes[0,1]}_{R\otimes Y} &
=  \hat U^{k_2} {\cal P}^{[r_1,r_2]\otimes[0,1]}_{R\otimes Y} \\
&&&\downarrow \\
&&&& \ldots \\
&&&&\downarrow \\
&&&&{\cal P}^{[r_s,r_{s-1}}_R &=  \hat U^{k_{s+1}} {\cal P}^{[r_s,1]}_R \\
&&&&&\downarrow \\
&&&&&{\cal P}^{[1,r_s]}_R &= {\cal M}_R
\label{iteschemelink}
\end{array}
\ee
There are two crucial differences: now the vertical arrows ($T$-duality) switches
between two different types of link polynomials: for the $B=[0,1]$ and $A=[1,0]$ unknots,
and these two types of link polynomials have two different evolutions
$\hat U$ and $\hat V$, which play the role of $\hat U^{\pm}$ in (\ref{ssev}).

{\bf The task is to define these evolutions and to explain why the vertical
arrows are now reversible.}
After one solves the problem for an arbitrary link polynomial of this type,
it is enough to put $Y=\emptyset$ to obtain the original torus polynomial.

The procedure is well defined and testable in the case of HOMFLY polynomials,
then it is just straightforwardly $t$-deformed.
It is well known that the result is not fully satisfactory:
some colored knot polynomials obtained in this way possess negative
coefficients and thus can not be considered as the superpolynomials of our dream. A way out
is supposed to be that what is obtained by this procedure is, in fact, the Euler characteristics of a
$t$-deformed complex, while there should be  corresponding Poincare polynomial of this complex, which
thus depends on an additional variable ${\bf t}$. Totally this gives as a polynomial of 4 variables $(A,q,t,{\bf t}$.
What is especially nice in (\ref{iteschemelink}),
it is clear, what is exactly assumed/postulated
about the $t$-deformation, and one can explicitly localize
the places which can be probably modified to improve the situation.

\section{Evolution on the topological locus}

Designing the evolution in the previous section implies a series of steps like
manifestly constructing the evolution operators and the seesaw mechanism which are quite non-trivial already at the level of
torus knots. This led us to the extended (off-shell) knot polynomials \cite{MMMkn1,MMMkn2}
which depend on an infinite set of variables,
but instead are no longer topological invariants
being associated not with the knot, but with its particular braid representation. These off-shell knot polynomials
become topological invariants only on the topological locus (\ref{tolo}) provided one multiplies them additionally
with a suitable monomial normalization factor \cite{IMMM2} which means choosing the proper (topological) framing.

\subsection{HOMFLY polynomials from evolution}

In practice, one can construct the evolution from the very beginning at the topological locus
generating the on-shell
HOMFLY polynomials much easier, for the price of impossibility to get the off-shell polynomials and, hence, of losing various
general structures related, in particular, to the double affine Hecke algebras \cite{Che}. To this end, we use the procedure
described in ss.\ref{21}: one determines the eigenvalues $\lambda_i$
of $R$-matrix which emerge after inserting "an evolution $m$-strand braid"
${\cal B}^n$ with $n$ repeating patterns inside the knot. This is done by decomposing the product of $m$ representations running
along the strands of $B$ into the sum of $p$ irreducible representations, so that each irreducible item gives rise to some
$\lambda_i$. In the general case, there are some coinciding eigenvalues so that there is a whole "eigenvalue matrix" $\Lambda_i$
in the space of intertwining operators, the result for the on-shell HOMFLY polynomial being the sum
\be
H_R^{{\cal E}{\cal B}^n}(A,q)=\sum_i^p\alpha_{R,i}(A,q)\Tr_{\cal M}\Lambda_i^n
\ee
where $\alpha_i$'s are some coefficients that depend on $q$ and $A$ but do not depend on $n$. Sometimes, however, all
representations enter the sum (\ref{tp}) with unit multiplicities so that all $\Lambda_i$'s are just numbers $\lambda_i$
\be\label{H}
H_R^{{\cal E}{\cal B}^n}(A,q)=\sum_i^p\alpha_{R,i}(A,q)\lambda_i^n
\ee
In this paper, we consider only such cases. Then, $\alpha_i$'s can be determined
from the $p$ known first terms of the infinite sequence of knots ${\cal K}^{(n)}$ associated with the evolution in $n$.
The eigenvalues $\lambda_i$ can be also functions of $q$ and $A$, but they are easily calculated. Though the $R$-matrices
often do not depend on $A$, in order to
obtain the topological framing, one has to normalize properly the ${\cal R}$-matrices and, hence, the eigenvalues, which introduces
an $A$-dependence anyway.

Moreover, this construction can be easily extended to include multiple evolutions: one suffices to consider a few series
of repeated patterns inserted into different parts of the knot. This would lead to the knot polynomial which depends on
a few evolution parameters $n_i$.

As we already mentioned in s.2, one has to differ between the evolution braids ${\cal B}^n$ made of "parallel" and
"anti-parallel" strands, since depending on the orientation one has to consider either representation $R$ or its
conjugate $\bar R$. In this paper, in order to avoid technical complications, and make the ideas of the evolution method as clear
as possible, we consider only two strand evolution braids. In this case,
the eigenvalue in the irreducible representation $Q$ of the "group" ${\cal R}$-matrix acting on the product of two
representations $R_1$ and $R_2$ are always $q^{\kappa_Q-\kappa_{R_1}-\kappa_{R_2}}$ (we do not care of the sign factors
$\epsilon_Q$, since they result only in common sign factors in $\alpha_i(A,q)$).
Thus, in the case of parallel strands they do not depend on $A$.
However, the group theory ${\cal R}$-matrices give rise to the vertical framing, which does not give the topological invariant
knot polynomials. In order to get the topological framing, one has to multiply the eigenvalues by the factor
$A^{-|Q|}q^{-\kappa_{R_1}-\kappa_{R_2}}$, where $|Q|$ denotes the size of the Young diagram $Q$. This immediately introduces an
$A$-dependence of the eigenvalues.

To be more concrete, let us consider symmetric representations $R=[r]$, which are our basic illustrative example in this paper.
Then, in the case of the parallel 2-strand braid the eigenvalues in the topological framing are:
\be\label{evp}
\lambda_i={q^{-r^2+i^2+i}\over A^r},\ \ \ \ \ \ \ i=0,\ldots,r
\ee
and correspond to the representations given by the two-line Young diagrams $Q=[r+i,r-i]$.
Similarly, in the case of the anti-parallel 2-strand braid the eigenvalues in the topological framing are
 (we use that $A=q^N$):
\be\label{evap}
\lambda_i=A^iq^{i(i-1)},\ \ \ \ \ \ \ i=0,\ldots,r
\ee
and correspond to the representations of $SU_q(N)$ given by the $N$-line Young diagrams $Q=[\underbrace{r+i,r,\ldots,r,r-i}_N]$.
The topological framing in this case of the anti-parallel 2-strand braid is determined by the requirement that the eigenvalue
of the singlet representation $i=0$, which is always present in the decomposition of $R\times\bar R$, is equal to unity.

\subsection{Superpolynomials from evolution}

Proceeding to superpolynomials,
we follow the suggestion of  \cite{DMMSS} and change the eigenvalues of the parallel braid in the topological framing,
\be
\lambda_Q = \frac{q^{\varkappa_Q}}{A^{|R|} q^{4\varkappa_R}} =
\frac{q^{\nu_{Q'}-\nu_Q}}{A^{|R|} q^{4\nu_{R'}-4\nu_R}}
\ \ \longrightarrow \ \
\frac{q^{\nu_{Q'}}t^{-\nu_Q}}{(Aq/t)^{|R|} q^{4\nu_{R'}}t^{-4\nu_R}}
\ee
Note that $A$ in the denominator is substituted by $Aq/t$. In particular, for the symmetric representation $R=[r]$
 (\ref{evp}) is replaced by
\be\label{sevp}
\lambda_i={t^iq^{-r^2+i^2}\over A^r},\ \ \ \ \ \ \ i=0,\ldots,r
\ee
Similarly,  (\ref{evap}) for the anti-parallel braid is replaced by  (we use that $A=t^N$)
\be\label{sevap}
\lambda_i=\left(A\cdot{q\over t}\right)^iq^{i(i-1)},\ \ \ \ \ \ \ i=0,\ldots,r
\ee
with the eigenvalues again normalized in such a way that the singlet representation corresponds to the unit eigenvalue.
In other words, in the first (parallel) case the non-deformed eigenvalue $\lambda_i$ is multiplies by $(t/q)^i$ and in the
second (anti-parallel) case it is divided by the same quantity.

Note that this substitution $A\to Aq/t$ works only for the eigenvalues, while the coefficients $\alpha_{R,i}(A,q)$
transform in a more tricky way, see examples below. The point is, however, that one suffices to know the eigenvalues, then
these coefficients can be {\bf evaluated} with the evolution method.

\section{Jones polynomials via Kauffman ${\cal R}$-matrix}

Before proceeding to more sophisticated applications of the evolution method,
to have solid grounds to stand on,
we first provide a sample calculation of the same quantities
within the $\mathcal{R}$-matrix approach.
We use its simplest version based on the Kauffman ${\cal R}$-matrix,
which provides descriptions of the Jones polynomials only, but instead
is very straightforward.
Our results below, obtained with the help of the more profound
evolution method generalize these simple formulas.

\subsection{Kauffman ${\cal R}$-matrix (fundamental representation of $SU_q(2)$)}

The $\mathcal{R}$-matrix ${\cal R}^{ij}_{kl}$ has four indices, which
in the fundamental representation of $SU_q(2)$
takes just two values $i,j,k,l=0,1$.
In this case one can also introduce two basic matrices of the
same kind:
\be
\|| = \delta^i_k\delta^j_l, \ \ \ \
\= = \epsilon^{ij}\epsilon_{kl}
\ee
In two dimensions the epsilon symbol is actually a canonical version
of the delta-symbol with two upper indices, and it flips the
orientation.
In terms of these basis matrices, the contraction of
indices is fully represented by connecting the lines, i.e.
all calculations are finally reduced to counting the number
of the closed cycles.
The agreement is that each closed cycle contributes
the quantum trace $D \equiv \Tr I = [2]_q = q+q^{-1}$.

The Kauffman $\mathcal{R}$-matrix \cite{Kaufm} and its inverse are
\be
{\cal R}(q) = q \|| - q^2\= , \nn \\
{\cal R}^{-1}(q) =\ q^{-1}\|| \ -\ q^{-2} \=   = {\cal R}(q^{-1})
\label{Kapict}
\ee

{
\begin{picture}(200,70)(-80,-30)
\put(-55,0){\mbox{${\cal R}^{ij}_{kl}$}}
\put(-32,0){\mbox{$=$}}
\put(-3,-3){\line(-1,-1){17}}
\put(3,3){\vector(1,1){17}}
\put(20,-20){\vector(-1,1){40}}
\put(-25,20){\mbox{$i$}}
\put(-25,-19){\mbox{$k$}}
\put(22,20){\mbox{$j$}}
\put(22,-19){\mbox{$l$}}
\put(28,0){\mbox{$=$}}
\put(40,-20){\vector(1,1){18}}
\put(80,-20){\vector(-1,1){18}}
\put(60,0){\vector(1,1){20}}
\put(60,0){\vector(-1,1){20}}
\put(60,0){\circle*{6}}
\put(88,0){\mbox{$=$}}
\put(110,0){\mbox{$q\ \cdot$}}
\qbezier(130,-20)(145,0)(130,20)
\qbezier(155,-20)(140,0)(155,20)
\put(137.3,0){\vector(0,1){2}}
\put(147.7,0){\vector(0,1){2}}
\put(165,0){\mbox{$\ \ -\ q^2\ \cdot$}}
%
\qbezier(200,20)(220,0)(240,20)
\qbezier(200,-15)(220,5)(240,-15)
\put(220,10){\circle*{4}}
\put(220,-5){\circle*{4}}
\put(217,-5.1){\vector(1,0){2}}
\put(223,-5.1){\vector(-1,0){2}}
\put(202,18){\vector(-1,1){2}}
\put(238,18){\vector(1,1){2}}
\put(258,0){\mbox{$= \ \ q\cdot\delta^i_k\delta^j_l
- q^2\cdot\epsilon^{ij}\epsilon_{kl}$}}
\end{picture}

\begin{picture}(200,70)(-80,-30)
\put(-75,0){\mbox{$\Big({\cal R}^{-1}\Big)^{ij}_{kl}$}}
\put(-32,0){\mbox{$=$}}
\put(-3,3){\vector(-1,1){17}}
\put(20,-20){\line(-1,1){17}}
\put(-20,-20){\vector(1,1){40}}
\put(-25,20){\mbox{$i$}}
\put(-25,-19){\mbox{$k$}}
\put(22,20){\mbox{$j$}}
\put(22,-19){\mbox{$l$}}
\put(28,0){\mbox{$=$}}
\put(40,-20){\vector(1,1){18}}
\put(80,-20){\vector(-1,1){18}}
\put(62,2){\vector(1,1){18}}
\put(58,2){\vector(-1,1){18}}
\put(60,0){\circle{6}}
\put(88,0){\mbox{$=$}}
\put(105,0){\mbox{$q^{-1}\ \cdot$}}
\qbezier(130,-20)(145,0)(130,20)
\qbezier(155,-20)(140,0)(155,20)
\put(137.3,0){\vector(0,1){2}}
\put(147.7,0){\vector(0,1){2}}
\put(165,0){\mbox{$-\  q^{-2}\ \cdot$}}
%
\qbezier(200,20)(220,0)(240,20)
\qbezier(200,-15)(220,5)(240,-15)
\put(220,10){\circle*{4}}
\put(220,-5){\circle*{4}}
\put(217,-5.1){\vector(1,0){2}}
\put(223,-5.1){\vector(-1,0){2}}
\put(202,18){\vector(-1,1){2}}
\put(238,18){\vector(1,1){2}}
\put(258,0){\mbox{$= \ \ q^{-1}\cdot\delta^i_k\delta^j_l
- q^{-2}\cdot\epsilon^{ij}\epsilon_{kl}$}}
\end{picture}
}

These two matrices do not change when both arrows are inverted,
therefore they can be combined (multiplied) both vertically and
horizontally, and the product is a unit matrix in both cases.
In fact, they respect {\bf all the three} Reidemeister moves
and the skein relation, see \cite{DM} for a detailed review.
Thus, they can be easily used for an {\it arbitrary} knot diagrams,
not restricted to the {\it braid} representations.

\bigskip

The twist ${\cal R}$-matrix and its inverse are:

\begin{picture}(200,70)(-50,-30)
\put(-55,0){\mbox{$\ho{\cal R}^{ij}_{kl}$}}
\put(-32,0){\mbox{$=$}}
\put(-3,3){\line(-1,1){17}}
\put(3,-3){\vector(1,-1){17}}
\put(-20,-20){\vector(1,1){40}}
\put(-25,20){\mbox{$i$}}
\put(-25,-19){\mbox{$k$}}
\put(22,20){\mbox{$j$}}
\put(22,-19){\mbox{$l$}}
\put(28,0){\mbox{$=$}}
\put(40,-20){\vector(1,1){18}}
\put(40,20){\vector(1,-1){18}}
\put(62,2){\vector(1,1){18}}
\put(62,-2){\vector(1,-1){18}}
\put(60,0){\circle*{6}}
\put(88,0){\mbox{$\!\!\!\!=$}}
\put(110,0){\mbox{$\!\!\!\!\!\!\!\!-q^{2}\ \cdot$}}
\qbezier(125,-20)(145,0)(125,20)
\qbezier(155,-20)(135,0)(155,20)
\put(135,0){\circle*{4}}
\put(145,0){\circle*{4}}
\put(135,-3){\vector(0,1){2}}
\put(135,3){\vector(0,-1){2}}
\put(153,18){\vector(1,1){2}}
\put(153,-18){\vector(1,-1){2}}
\put(165,0){\mbox{$ +\ q^{}\ \cdot\!\!$}}
\qbezier(200,20)(220,0)(240,20)
\qbezier(200,-15)(220,5)(240,-15)
\put(238,-13){\vector(1,-1){2}}
\put(238,18){\vector(1,1){2}}
\put(258,0){\mbox{$= \ \ -q^{2}\cdot\epsilon^{jl}\epsilon_{ik}
+ q\cdot\delta^j_i\delta^l_k = {\cal R}^{jl}_{ik}$}}
\end{picture}

\begin{picture}(200,70)(-50,-30)
\put(-70,0){\mbox{$\Big(\ho{\cal R}^{-1}\Big)^{ij}_{kl}$}}
\put(-32,0){\mbox{$=$}}
\put(-3,-3){\line(-1,-1){17}}
\put(3,3){\vector(1,1){17}}
\put(-20,20){\vector(1,-1){40}}
\put(-25,20){\mbox{$i$}}
\put(-25,-19){\mbox{$k$}}
\put(22,20){\mbox{$j$}}
\put(22,-19){\mbox{$l$}}
\put(28,0){\mbox{$=$}}
\put(40,-20){\vector(1,1){18}}
\put(40,20){\vector(1,-1){18}}
\put(62,2){\vector(1,1){18}}
\put(62,-2){\vector(1,-1){18}}
\put(60,0){\circle{6}}
\put(88,0){\mbox{$\!\!\!\!=$}}
\put(110,0){\mbox{$\!\!\!\!\!\!\!\!-q^{-2}\ \cdot$}}
\qbezier(125,-20)(145,0)(125,20)
\qbezier(155,-20)(135,0)(155,20)
\put(135,0){\circle*{4}}
\put(145,0){\circle*{4}}
\put(135,-3){\vector(0,1){2}}
\put(135,3){\vector(0,-1){2}}
\put(153,18){\vector(1,1){2}}
\put(153,-18){\vector(1,-1){2}}
\put(165,0){\mbox{$ +\ q^{-1}\ \cdot\!\!$}}
\qbezier(200,20)(220,0)(240,20)
\qbezier(200,-15)(220,5)(240,-15)
\put(238,-13){\vector(1,-1){2}}
\put(238,18){\vector(1,1){2}}
\put(258,0){\mbox{$= \ \ -q^{-2}\cdot\epsilon^{jl}\epsilon_{ik}
+ q^{-1}\cdot\delta^j_i\delta^l_k = \Big({\cal R}^{-1}\Big)^{jl}_{ik}$}}
\end{picture}

However, as we already mentioned, in practice there is no difference
between $\delta$ and $\epsilon$ symbols, what matters at the end
is just the number of closed contours.
This means that from calculational point of view one can
actually substitute
\be
\ho{\cal R} \cong -q^{3}{\cal R}^{-1},\nn \\
\ho{\cal R}^{-1} \cong -q^{-3} {\cal R}
\ee
This fact can be described from the point of view of eigenvalues of $\mathcal{R}$-matrices.
It corresponds to the identities $-1\cong -q^3A^{-1}q^{-1}$ and $A \cong -q^3A^{-1}(-q)$ between the eigenvalues
of $\bar{\cal R}$ and ${\cal R}^{-1}$ (see below),
which are indeed true for $A=q^2$.
A related remark is that one can apply the first Reidemeister move in the vertical channel:
${\cal R}^{ij}_{kj} = \delta^i_j$ and ${\cal R}^{ij}_{il} = \delta^j_l$, but in the horizontal channel one has instead
${\cal R}^{ij}_{kl}\epsilon^{kl} = -q^3\epsilon^{ij}$ and
${\cal R}^{ij}_{kl}\epsilon_{ij} = -q^3\epsilon_{kl}$.

The projectors onto representations $[2]$ and $[11]$
(the latter is indistinguishable from the singlet at the $SU_q(2)$ level)
in the vertical direction are
\be
P_{[11]}^{vert} = \frac{1}{D}\=
\ \ \ \ \ \ \ {\rm and}\ \ \ \ \ \ \
P_{[2]}^{vert} = I^{vert} - P_{[11]}^{vert} = \ \|| - \ \frac{1}{D}\=
\ee
Similarly, the projectors in the horizonal direction are:
\be
P_{[11]}^{hor} = \frac{1}{D}\||
\ \ \ \ \ \ \ {\rm and}\ \ \ \ \ \ \
P_{[2]}^{vert} = I^{hor} - P_{[11]}^{hor} = \ \=  - \ \frac{1}{D}\||
\ee

Using these ingredients it is easy to develop a kind of "a
Reggeon calculus" for the Jones polynomials of knots,
consisting of two kinds of the $2$-strand braids, parallel and anti-parallel.
It can be then further extended to $m$-strand braids
and/or to the colored Jones polynomials.

\subsection{$2$-strand braids in the fundamental representation}

\subsubsection{Chains of ${\cal R}$-matrices}

First of all, one can consider the chains of the ${\cal R}$-matrices
(\ref{Kapict}).
They can form vertical and horizontal chains:

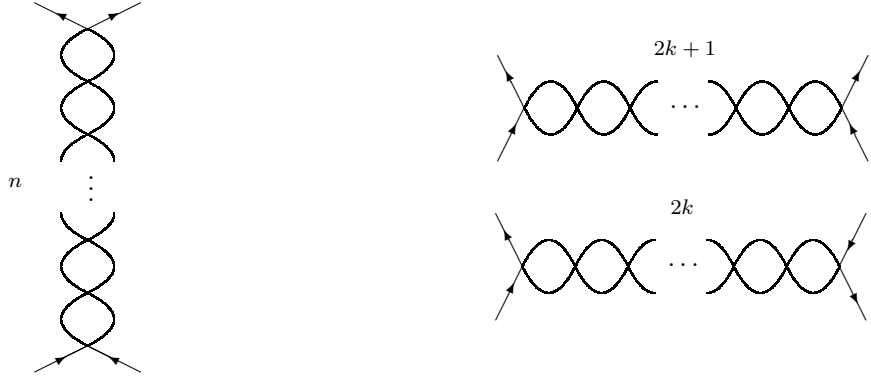
\begin{picture}(100,170)(-100,-220)
\qbezier(0,-80)(-20,-90)(0,-100)
\qbezier(0,-80)(20,-90)(0,-100)
\qbezier(0,-100)(-20,-110)(0,-120)
\qbezier(0,-100)(20,-110)(0,-120)
\qbezier(0,-120)(-10,-125)(-10,-130)
\qbezier(0,-120)(10,-125)(10,-130)
\put(0,-145){\mbox{$\vdots$}}
\qbezier(0,-160)(-10,-155)(-10,-150)
\qbezier(0,-160)(10,-155)(10,-150)
\qbezier(0,-160)(-20,-170)(0,-180)
\qbezier(0,-160)(20,-170)(0,-180)
\qbezier(0,-180)(-20,-190)(0,-200)
\qbezier(0,-180)(20,-190)(0,-200)
\put(0,-80){\line(-2,1){20}}
\put(0,-80){\line(2,1){20}}
\put(0,-200){\line(-2,-1){10}}
\put(0,-200){\line(2,-1){10}}
\put(0,-200){\line(-2,-1){20}}
\put(0,-200){\line(2,-1){20}}
\put(-10,-205){\vector(2,1){2}}
\put(10,-205){\vector(-2,1){2}}
\put(-10,-75){\vector(-2,1){2}}
\put(10,-75){\vector(2,1){2}}
\put(-30,-140){\mbox{\footnotesize $n$}}
\end{picture}
\begin{picture}(200,50)(-223,-110)
\qbezier(-40,0)(-50,20)(-60,0)
\qbezier(-40,0)(-50,-20)(-60,0)
\qbezier(-20,0)(-30,20)(-40,0)
\qbezier(-20,0)(-30,-20)(-40,0)
\qbezier(-20,0)(-15,10)(-10,10)
\qbezier(-20,0)(-15,-10)(-10,-10)
\put(-5,0){\mbox{$\ldots$}}
\qbezier(10,10)(15,10)(20,0)
\qbezier(10,-10)(15,-10)(20,0)
\qbezier(20,0)(30,20)(40,0)
\qbezier(20,0)(30,-20)(40,0)
\qbezier(40,0)(50,20)(60,0)
\qbezier(40,0)(50,-20)(60,0)
\put(-60,0){\line(-1,2){10}}
\put(-60,0){\line(-1,-2){10}}
\put(60,0){\line(1,2){10}}
\put(60,0){\line(1,-2){10}}
\put(-65,-10){\vector(1,2){2}}
\put(-65,10){\vector(-1,2){2}}
\put(65,10){\vector(1,2){2}}
\put(65,-10){\vector(-1,2){2}}
\put(-11,20){\mbox{\footnotesize$2k+1$}}
\end{picture}
\begin{picture}(200,50)(-20,-50)
\qbezier(-40,0)(-50,20)(-60,0)
\qbezier(-40,0)(-50,-20)(-60,0)
\qbezier(-20,0)(-30,20)(-40,0)
\qbezier(-20,0)(-30,-20)(-40,0)
\qbezier(-20,0)(-15,10)(-10,10)
\qbezier(-20,0)(-15,-10)(-10,-10)
\put(-5,0){\mbox{$\ldots$}}
\qbezier(10,10)(15,10)(20,0)
\qbezier(10,-10)(15,-10)(20,0)
\qbezier(20,0)(30,20)(40,0)
\qbezier(20,0)(30,-20)(40,0)
\qbezier(40,0)(50,20)(60,0)
\qbezier(40,0)(50,-20)(60,0)
\put(-60,0){\line(-1,2){10}}
\put(-60,0){\line(-1,-2){10}}
\put(60,0){\line(1,2){10}}
\put(60,0){\line(1,-2){10}}
\put(-65,-10){\vector(1,2){2}}
\put(-65,10){\vector(-1,2){2}}
\put(65,10){\vector(-1,-2){2}}
\put(65,-10){\vector(1,-2){2}}
\put(-4,20){\mbox{\footnotesize$2k$}}
\end{picture}

\bigskip

Substituting the Kauffman $\mathcal{R}$-matrix into the vertical chain gives:
$$
a_{n+1} \|| + b_{n+1} \= \ \ =
\ \ \Big\{\ a_n \|| + b_n \=\Big\} \cdot_{\rm vert} \Big\{\ q\|| -q^2 \=\Big\}\ \
=\ \  qa_n \||  + \Big(qb_n-q^2a_n -q^2b_nD\Big)\cdot\=
$$
\vspace{-0.4cm}
\be
\Longrightarrow \ \
a_{n+1} = qa^n,\ \ \ \ b_{n+1} = -q^2a_n - q^3b_n
\ee
and it follows from this recurrent formula that
\be
\boxed{
B^{[2,n]}_{_{\Box,N=2}} =
q^n \left\{ \|| \ - \frac{1}{D}\Big(1-(-)^nq^{2n}\Big)\cdot\= \right\}
}
\label{B2n}
\ee

Substituting the same relations into the horizontal chain ($n=2k$ or $n=2k+1$) gives:
$$
\ho{a}_{n+1} \|| + \ho{b}_{n+1} \= =
\left\{\ \ho{a}_n \|| + \ho{b}_n\=\right\}\cdot_{hor}
\left\{\ q\|| - q^2\=\right\}
$$
\vspace{-0.4cm}
\be
\Longrightarrow\ \
\ho{b}_{n+1} = -q^2\ho{b}_n,\ \ \ \ \ \
\ho{a}_{n+1} = \ho{a}_n + q\ho{b}_n
\ee
and the result is
\be
\boxed{
\ho{B}^{[2,n]}_{_{\Box,N=2}}
= \frac{1}{D}\Big(1-(-q^2)^n\Big)\cdot\|| \ + (-q^2)^n \=
}
\label{hoB2n}
\ee

The same can be done with the twisted matrices $\ho{\cal R}$.
The difference is in the direction of arrows:
the vertical (horizontal) chain of $\ho{\cal R}$ is
equivalent to the horizontal (vertical) chain of ${\cal R}$,
but one should also exchange $\=$ with $\||$ in the answer.
In what follows we call this operation transposition of braids.

\subsubsection{Effective vertices. Valence $2$}

"Effective vertices", describing "interactions"
of our "Reggeons", are "boxes" with $2k$ external legs,
$k$ incoming and $k$ outgoing,
and they can have different "internal" structure.

The simplest is the vertex of valence $2k=2$  with no internal structure,
just converting the $jl$ or $ik$ indices of ${\cal R}^{ij}_{kl}$
or the $ij$ or $kl$ indices of $\overline{{\cal R}}^{ij}_{kl}$.
Since the Kauffman ${\cal R}$-matrix respects the first Reidemeister move,
this vertex simply ``annihilates'' the ${\cal R}$-matrix.
One can apply this operation twice to (\ref{B2n}) or to (\ref{hoB2n})
in the horizontal and vertical channels,
gluing respectively the two ends at the same height
or when one is above the other, i.e. by the rule

\begin{picture}(200,80)(-120,-40)
\put(0,0){\circle{40}}
\qbezier(-30,-30)(0,50)(30,-30)
\put(-28,-25){\vector(1,2){2}}
\put(26,-20){\vector(1,-2){2}}
\put(60,0){\circle{40}}
\qbezier(30,30)(60,-50)(90,30)
\put(34,20){\vector(-1,2){2}}
\put(88.5,25){\vector(-1,-2){2}}
\put(25,-50){\mbox{hor}}
\put(200,0){\circle{40}}
\qbezier(170,-30)(250,0)(170,30)
\put(178,27){\vector(2,-1){2}}
\put(178,-27){\vector(-2,-1){2}}
\put(260,0){\circle{40}}
\qbezier(290,-30)(210,0)(290,30)
\put(283,27.3){\vector(-2,-1){2}}
\put(283,-27){\vector(2,-1){2}}
\put(225,-50){\mbox{vert}}
\end{picture}

\be
\begin{array}{ccc}
hor:  &  \|| \longrightarrow D & \ \ \ \ \= \longrightarrow D^2 \cr
&&\cr
vert: &  \|| \longrightarrow D^2 & \ \ \ \ \= \longrightarrow D
\label{val2op}
\end{array}
\ee
The vertical gluing converts the vertical braid into a torus knot/link,
while the horizontal gluing (it actually involves the $\epsilon$ tensors
in this case, but this does not matter for our consideration)
into the unknot.
Indeed, applying (\ref{val2op}) to (\ref{B2n}) one gets
\be
\begin{array}{ccc}
hor: & B_{_{\Box,N=2}}^{[2,n]} \longrightarrow &  (-)^nq^{3n} D
= (-)^nq^{3n} J_{_\Box}^{\rm unknot} \cr
&&\cr
vert: & B_{_{\Box,N=2}}^{[2,n]} \longrightarrow &
q^n\Big(q^{-2}+1+q^2+(-)^nq^{2n}\Big) = J_{_\Box}^{[2,n]}(q)
\end{array}
\ee
Vice versa, the horizontal braid is converted into the unknot by the vertical gluing,
while the horizontal gluing makes it into a torus knot/link,
and application of (\ref{val2op}) to (\ref{hoB2n}) gives
\be
\begin{array}{ccc}
hor: & \ho{B}_{_{\Box,N=2}}^{[2,n]} \longrightarrow &
1 + (-)^nq^{2n}(q^{-2}+1+q^2) =  \left(-q^3\right)^n J_{_\Box}^{[2,n]}(q^{-1}) \cr
&&\cr
vert: & \ho{B}_{_{\Box,N=2}}^{[2,n]} \longrightarrow &
D =  J_{_\Box}^{\rm unknot}
\end{array}
\ee

\subsubsection{Effective vertices of valence $4$: twist knots}

Besides the valence $2$ vertices one can also look at the valence $4$ vertices.
There are of course infinitely many of them. The simplest one is just equal to the combination of two valence
$2$ vertices and was already studied in the previous subsection. The next to the simplest is the vertex consisting of
one crossing. This transforms the chain of the $\mathcal{R}$-matrices into the chain with one more matrix and thus
also is not very interesting.

The simplest nontrivial valence $4$ vertex is the twist blocks. They can be either
\be
{\rm T}=\Big\{q\|| - q^2\=\Big\}\cdot_{\rm hor}\Big\{q\||-q^2\=\Big\}\ =\
(q^2D-2q^3)\|| + q^4\cdot\= =
q^2(q^{-1}-q)\cdot\|| + q^4\= =
\ee
or
\be
\Big\{q^{-1}\|| -q^{-2}\=\Big\}\cdot_{\rm hor}\Big\{q^{-1}\|| -q^{-2}\=  \Big\}
\ =\ (q^{-2}D-2q^{-3})\|| + q^{-4}\cdot\= \ =\
- q^{-3}(q-q^{-1})\cdot\|| + q^{-4}\=
\nn
\ee
Of course, these formulas are particular cases of (\ref{hoB2n}) for
$n=2$ and $n=-2$ respectively.

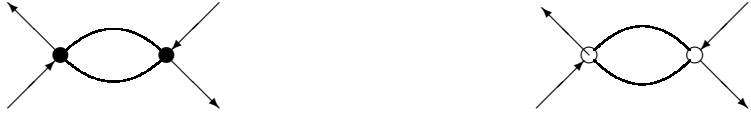
\begin{picture}(200,70)(-100,-35)
\put(0,0){\vector(-1,1){20}}
\put(-20,-20){\vector(1,1){18}}
\put(0,0){\circle*{6}}
\put(40,0){\circle*{6}}
\put(60,20){\vector(-1,-1){18}}
\put(40,0){\vector(1,-1){20}}
\qbezier(0,0)(20,20)(40,0)
\qbezier(0,0)(20,-20)(40,0)
\put(200,0){\vector(-1,1){18}}
\put(180,-20){\vector(1,1){18}}
\put(200,0){\circle{6}}
\put(240,0){\circle{6}}
\put(260,20){\vector(-1,-1){18}}
\put(242,-2){\vector(1,-1){18}}
\qbezier(202,2)(220,20)(238,2)
\qbezier(202,-2)(220,-20)(238,-2)
\end{picture}

Projecting this block onto two irreducible representations
in the vertical channel, one gets
\be
{\rm T}_{[11]} = \frac{1}{D} \Big(q^4D +q^2(q^{-1}-q^1)\Big)\=
\ =\ q^3\frac{q^2+q^{-2}}{q+q^{-1}}\= = q^3(q^2+q^{-2})P_{[11]}, \nn \\
{\rm T}_{[2]} = -q^2(q-q^{-1})\|| + q^4\= - {\rm T}_{[11]}
= -q^2(q-q^{-1})P_{[2]}
\ee

To this one should add the vertical trace of the two projectors:
\be
\Tr_{vert} P_{[11]} = 1, \ \ \ \ \
\Tr_{vert} P_{[2]} = D^2-1 = q^2+1+q^{-2} = \frac{\{q^3\}}{\{q\}}
\ee
Then
\be
\frac{1}{D}\Tr_{vert} {\rm T}_{[11]} = \frac{q^3(q^2+q^{-2})}{q+q^{-1}}
= \left.\frac{A^2(q^2-1+q^{-2}-A^2)}{1-A^2}\right|_{A=q^2}, \nn \\
\frac{1}{D}\Tr_{vert} {\rm T}_{[2]} = -\frac{q^2(q^3-q^{-3})}{q+q^{-1}}
= -\left.\frac{A \{Aq\}\{A/q\}}{\{A\}}\right|_{A=q^2}
\ee
reproduce the values of the coefficients
$\alpha_{[1],i}$ in  (\ref{atw1}) at $A=q^2$.

One can now attach to the transposition of this block the horizontal braid
(\ref{hoB2n}) to obtain the twisted knot. Only the braid with the even number of crossings should be used
which become obvious if one tries to draw the diagram of the knot. For the odd crossings braid, the block
with the orientation of one strand inverted should be used. But such an odd crossing knot can be easily transformed
into an even number knot with one less crossing. The answer for the braid with $2n$ crossings, or for the $n$-twist knot:
\be
\widetilde{T}\cdot \ho{B}^{[2,2n]}_{_{\Box,N=2}}
=
\Big\{\ q^4\|| -q^2(q-q^{-1})\cdot\=\Big\}\cdot_{hor}
\left\{\frac{1}{D}\Big(1-q^{4n}\Big)\cdot\|| \ + q^{4n} \=\right\}
= \nn \\ =
\left\{q^4(1-q^{4n})-q^2\Big(q-q^{-1}\Big)\frac{1}{D}\Big(1-q^{4n}\Big)+q^{4n+4}
\right\}\||
- (q-q^{-1})q^{4n+2}\= \ = \nn \\ =
\frac{1}{D}\left(q^{4n+3}-q^{4n+1}+q^5+q\right)\||
- (q-q^{-1})q^{4n+2}\= \nn \\
\Longrightarrow \ \
J^{T^{(n)}}_{_\Box} =
\Big(q^{4n+3}-q^{4n+1}+q^5+q - (q-q^{-1})q^{4n+2}D^2\Big)
=D\Big(q^{4n}+q^4-q^{4n+4}+(q^3-q)\frac{q^{4n}-1}{D}\Big)
\ee
which is proportional to (\ref{felf}) at $A=q^2$ up to the normalization factor $D$, since the Jones polynomial is
unreduced here.

\subsubsection{Effective vertices of valence $4$: double braids\label{doubleb}}

The next in simplicity valence $4$ vertex after the twist blocks are the $\mathcal{R}$-matrix chains discussed above.
The obvious generalization of these chains and thus the torus and twist links and knots is the double braids.
There are two families of such braids:

\bigskip

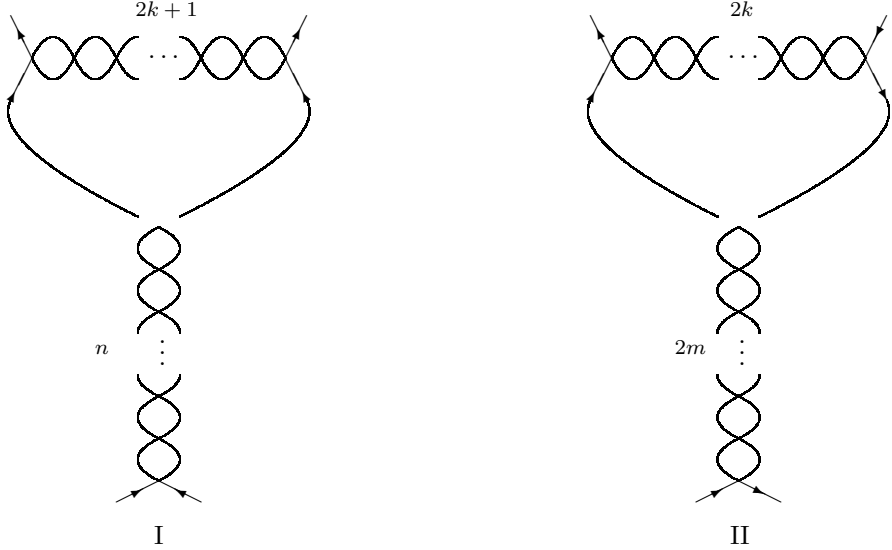
\unitlength=0.8pt
\begin{picture}(200,270)(-100,-240)
\qbezier(-40,0)(-50,20)(-60,0)
\qbezier(-40,0)(-50,-20)(-60,0)
\qbezier(-20,0)(-30,20)(-40,0)
\qbezier(-20,0)(-30,-20)(-40,0)
\qbezier(-20,0)(-15,10)(-10,10)
\qbezier(-20,0)(-15,-10)(-10,-10)
\put(-5,0){\mbox{$\ldots$}}
\qbezier(10,10)(15,10)(20,0)
\qbezier(10,-10)(15,-10)(20,0)
\qbezier(20,0)(30,20)(40,0)
\qbezier(20,0)(30,-20)(40,0)
\qbezier(40,0)(50,20)(60,0)
\qbezier(40,0)(50,-20)(60,0)
\put(-60,0){\line(-1,2){10}}
\put(-60,0){\line(-1,-2){10}}
\put(60,0){\line(1,2){10}}
\put(60,0){\line(1,-2){10}}
\qbezier(0,-80)(-20,-90)(0,-100)
\qbezier(0,-80)(20,-90)(0,-100)
\qbezier(0,-100)(-20,-110)(0,-120)
\qbezier(0,-100)(20,-110)(0,-120)
\qbezier(0,-120)(-10,-125)(-10,-130)
\qbezier(0,-120)(10,-125)(10,-130)
\put(0,-145){\mbox{$\vdots$}}
\qbezier(0,-160)(-10,-155)(-10,-150)
\qbezier(0,-160)(10,-155)(10,-150)
\qbezier(0,-160)(-20,-170)(0,-180)
\qbezier(0,-160)(20,-170)(0,-180)
\qbezier(0,-180)(-20,-190)(0,-200)
\qbezier(0,-180)(20,-190)(0,-200)
\put(0,-80){\line(-2,1){10}}
\put(0,-80){\line(2,1){10}}
\put(0,-200){\line(-2,-1){10}}
\put(0,-200){\line(2,-1){10}}
\put(0,-200){\line(-2,-1){20}}
\put(0,-200){\line(2,-1){20}}
\qbezier(-10,-75)(-80,-40)(-70,-20)
\qbezier(10,-75)(80,-40)(70,-20)
\put(-10,-205){\vector(2,1){2}}
\put(10,-205){\vector(-2,1){2}}
\put(-65,10){\vector(-1,2){2}}
\put(65,10){\vector(1,2){2}}
\put(-70,-20){\vector(1,2){2}}
\put(70,-20){\vector(-1,2){2}}
\put(-11,20){\mbox{\footnotesize$2k+1$}}
\put(-30,-140){\mbox{\footnotesize $n$}}
\put(-2,-230){\mbox{I}}
\end{picture}
\begin{picture}(200,270)(-170,-240)
\qbezier(-40,0)(-50,20)(-60,0)
\qbezier(-40,0)(-50,-20)(-60,0)
\qbezier(-20,0)(-30,20)(-40,0)
\qbezier(-20,0)(-30,-20)(-40,0)
\qbezier(-20,0)(-15,10)(-10,10)
\qbezier(-20,0)(-15,-10)(-10,-10)
\put(-5,0){\mbox{$\ldots$}}
\qbezier(10,10)(15,10)(20,0)
\qbezier(10,-10)(15,-10)(20,0)
\qbezier(20,0)(30,20)(40,0)
\qbezier(20,0)(30,-20)(40,0)
\qbezier(40,0)(50,20)(60,0)
\qbezier(40,0)(50,-20)(60,0)
\put(-60,0){\line(-1,2){10}}
\put(-60,0){\line(-1,-2){10}}
\put(60,0){\line(1,2){10}}
\put(60,0){\line(1,-2){10}}
\qbezier(0,-80)(-20,-90)(0,-100)
\qbezier(0,-80)(20,-90)(0,-100)
\qbezier(0,-100)(-20,-110)(0,-120)
\qbezier(0,-100)(20,-110)(0,-120)
\qbezier(0,-120)(-10,-125)(-10,-130)
\qbezier(0,-120)(10,-125)(10,-130)
\put(0,-145){\mbox{$\vdots$}}
\qbezier(0,-160)(-10,-155)(-10,-150)
\qbezier(0,-160)(10,-155)(10,-150)
\qbezier(0,-160)(-20,-170)(0,-180)
\qbezier(0,-160)(20,-170)(0,-180)
\qbezier(0,-180)(-20,-190)(0,-200)
\qbezier(0,-180)(20,-190)(0,-200)
\put(0,-80){\line(-2,1){10}}
\put(0,-80){\line(2,1){10}}
\put(0,-200){\line(-2,-1){10}}
\put(0,-200){\line(2,-1){10}}
\put(0,-200){\line(-2,-1){20}}
\put(0,-200){\line(2,-1){20}}
\qbezier(-10,-75)(-80,-40)(-70,-20)
\qbezier(10,-75)(80,-40)(70,-20)
\put(-10,-205){\vector(2,1){2}}
\put(10,-205){\vector(2,-1){2}}
\put(-65,10){\vector(-1,2){2}}
\put(67,14){\vector(-1,-2){2}}
\put(-70,-20){\vector(1,2){2}}
\put(68,-16){\vector(1,-2){2}}
\put(-4,20){\mbox{\footnotesize$2k$}}
\put(-30,-140){\mbox{\footnotesize $2m$}}
\put(-4,-230){\mbox{II}}
\end{picture}

Case I corresponds to the torus knot or link $[2,n+1]$ when $k=0$ and case II corresponds to the twist knot $T^{(m)}$
when $k=-1$.

The answer for the fundamental Jones polynomial of the knot or link (case I),
where $2k+1$ is the number of crossings in the twist braid
and $n$ is the number of crossings in the torus braid, is
\be\label{Jonesmtontw}
D\left(q^{-n}+q^{4k+2-n}-(-1)^n q^{4k+2-3n}\right)
+\frac{1}{D}\left((-q^{-3})^n(1+q^{4k+2})-q^{-n}-q^{4k+2-n}\right)
= \nn \\
= q^{-n}\left\{ D\Big(1+q^{4k+2} - (-q^2)^{-n}q^{4k+2}\Big)
-  \frac{1+q^{4k+2}}{D}\Big(1-(-q^2)^{-n}\Big)\right\}
\ee
In the second term, $1+q^{4k+2}$ is always divisible by $D$.

The answer for the two twist braids (case II) $2n$ and $2m$ is:
\be
D\left(q^{4n}+q^{-4m}-q^{4n-4m}\right)+\frac{1}{D}\left(1-q^{4n}-q^{-4m}+q^{4n-4m}
\right)
\ee

\section{Applying the evolution method}

In this section we illustrate the technique explained in section 3 with simplest examples of knot series containing
parallel and anti-parallel 2-strand braids: the torus 2-strand knots (the parallel strands), the twist knots (anti-parallel
strands), the 2-strand links with opposite orientation of the components (anti-parallel
strands) and the double braid (double evolution) knots (containing both the parallel and anti-parallel strands). While
the first three cases are well-known, though we suggest their different interpretation, the results on the double braid knots,
to our best of knowledge, have been not presented in the literature so far.

\subsection{Torus knots\label{torus}}

We start with the series of (torus) knots $T^{(2,2k+1)}$ that are described by the two parallel strands with
$2k+1$ intersections (when the number of intersections is even, it is a two-component link)
and calculate their HOMFLY polynomials
$H_R^{T^{(2,2k+1)}}(A,q)=X^k_R(A,q)$ by the evolution method. We already partly discussed this example in ss.\ref{21}.

\subsubsection{The fundamental representation $R=[1]$}

In this case there are two eigenvalues
\be
\lambda_0={1\over qA}\ \ \ \ \  \ \ \ \hbox{and}\ \ \ \ \ \ \ \lambda_1={q\over A}
\ee
so that the general form of the polynomial looks like (\ref{2}), and the answer is (see s.2.1)
\be\label{torusf}
\alpha_{[1],0}=-{\{A/q\}\over \{q^2\}},\ \ \ \ \ \ \ \ \ \alpha_{[1],1}= {\{Aq\}\over\{q^2\}}\nn\\
X^{(k)}_{[1]}(A,q)={\{Aq\}q^{4k+2}-\{A/q\}\over \{q^2\}(qA)^{2k+1}}
\ee
In particular, for the trefoil one gets
$X^{(-2)}_{[1]}(q,A)=A^2(q^2+1/q^2)-A^4$, which is correct.
The answer can be also rewritten in the form
\be\label{felft}
X^{(k)}_{[1]}(A,q)=1 - A^{-2} h_k(A^2|q^2)\{Aq\}\{A/q\}
\ee
with
\be
h_1 = 1, \nn \\
h_2 = 1+ (q^2+q^{-2})A^{-2}, \nn \\
h_3 = 1+ (q^2+q^{-2})A^{-2} + (q^4+1+q^{-4})A^{-4}, \nn \\
\ldots \nn \\
h_r-h_{r-1} = \frac{q^{2r}-q^{-2r}}{q^2-q^{-2}}A^{2-2r}
\ee
Note that
\be\label{sym}
X^{(k)}_{[1]}(A,q)=X^{(-k-1)}_{[1]}(A^{-1},q^{-1})
\ee

\subsubsection{Symmetric representations}

Similarly, for the first symmetric representation there are three
eigenvalues (\ref{evp}) and, hence, three unknown $\alpha_i$'s which can be obtained from the values of the unknot and
the trefoil in the first symmetric representation, the result reads (see (\ref{12}))
\be
X^{(k)}_{[2]}(A,q)={\{A\}\{A/q\}\over\{q^2\}\{q^3\}}\left({1\over A^2q^4}\right)^{2k+1}
-{\{A/q\}\{Aq^2\}\over\{q\}\{q^4\}}\left({1\over A^2q^2}\right)^{2k+1}+{\{Aq^2\}\{Aq^3\}\over
\{q^3\}\{q^4\}}\left({q^2\over A^2}\right)^{2k+1}
\ee
Analogously for representation $[3]$ one obtains
\be
X^{(k)}_{[3]}(A,q)=-\{q\}\cdot{\{A/q\}\{A\}\{Aq\}\over\{q\}\{q^2\}\{q^3\}\cdot\{q^4\}}\left({1\over A^3q^9}\right)^{2k+1}
+\{q^3\}\cdot{\{Aq^3\}\cdot\{A/q\}\{A\}\over\{q\}\{q^2\}\cdot\{q^4\}\{q^5\}}\left({1\over A^3q^7}\right)^{2k+1}-\\
-\{q^5\}\cdot{\{Aq^3\}\{Aq^4\}\cdot\{A/q\}\over\{q\}\cdot\{q^4\}\{q^5\}\{q^6\}}\left({1\over A^3q^3}\right)^{2k+1}
+\{q^7\}\cdot{\{Aq^3\}\{Aq^4\}\{Aq^5\}\over\{q^4\}\{q^5\}\{q^6\}\{q^7\}}\left({q^3\over A^3}\right)^{2k+1}
\ee
These examples are already enough to guess the general formula:
\be
\boxed{
X^{(k)}_{[r]}(A,q) = \sum_{l=0}^r (-)^{r+l} \{q^{2l+1}\}
\frac{\prod_{i=0}^{l-1} \{Aq^{r+i}\} \prod_{i=0}^{r-l-1} \{Aq^{i-1}\}}
{\prod_{i=1}^{r-l} \{q^i\} \prod_{i=r+1}^{r+l+1} \{q^i\}} \left[ {q^{\ell(\ell+1)-r^2}\over A^r} \right]^{2k+1}
}
\ee
This coincides with the suggestion of \cite{FGS1}
in terms of quantum Pochhammer symbols
\be
X^{(k)}_{[r]}(A,q)=\sum_{\ell=0}^r (-1)^{r+\ell}
\frac{(q)_r(A)_{r+\ell}(A/q)_{r-\ell}}
{(A)_r(q^2)_{r+\ell}(q)_{r-\ell}}
 \frac{\{{q}^{2\ell+1}\}}{\{q\}}
\left[ {q^{\ell(\ell+1)-r^2}\over A^r} \right]^{2k+1}
\ee
Note that the symmetry (\ref{sym}) still persists in the generic case:
\be\label{symr}
X^{(k)}_{[r]}(A,q)=X^{(-k-1)}_{[r]}(A^{-1},q^{-1})
\ee

\subsubsection{Superpolynomial in symmetric representation $R=[r]$}

Using formulas (\ref{sevp}), one can obtain in the fundamental case for the superpolynomials
of the 2-strand torus knots ($n=2k+1$)
\be
P_{\Box}^{T[2,n]} (A,q,t)= \left(\frac{t}{q}\right)^n\cdot
\left(\frac{\{Aq\}}{\{qt\}}\left(\frac{q}{A}\right)^n
- \frac{\{A/t\}}{\{qt\}}\left(\frac{1}{At}\right)^n\right)
\ee
This expression can be alternatively rewritten as
\be
P_{\Box}^{T[2,n]}(A,q,t) =  \frac{\{Aq\}}{\{qt\}}\left(\frac{t}{A}\right)^n
- \frac{\{A/t\}}{\{qt\}}\left(\frac{1}{Aq}\right)^n
\ee
(which looks like made from $t^{\nu_{Q'}}q^{-\nu_Q}$ instead of
$q^{\nu_{Q'}}t^{-\nu_Q}$),
but there is no such representation for $R\neq \Box$.
Still, in this form it is clear that the eigenvalues $t/A$ and $1/Aq$
are obtained from the HOMFLY ones, $q/A$ and $1/Aq$ by the change
of variables $(A^2,q^2)\longrightarrow (A^2q/t, qt)$.
Since the initial conditions $P_{_\Box}^{[2,\pm 1]} = 1$ are not $t$-deformed
at all, this means that the whole answer in this case
is obtained by a change of variables from HOMFLY:
\be
P_{_\Box}^{[2,n]}(A,q,t)
=  1 +\, G_1^{(n)}\!\left(\frac{A^2q}{t},qt\right)\{Aq\}\{A/t\}
\ee

This is not just so simple for higher representations,
already for $R=[2]$:
\be
P_{[2]}^{[2,n]}(A,q,t) = \frac{\{Aq^3\}\{Aq^2\}}{\{q^3t\}\{q^2t\}}\left(\frac{t^2}{A^2}\right)^n
-(q+q^{-1}) \frac{\{Aq^2\}\{A/t\}}{\{q^3t\}\{qt\}}\left(\frac{t}{A^2q^3}\right)^n
+ \frac{\{Aq/t\}\{A/t\}}{\{q^2t\}\{qt\}}\left(\frac{1}{A^2q^4}\right)^n
\ee
Note that the coefficient in the second term contains the ratio
$(q+q^{-1})/\{qt\}$ which turns into $1/\{q\}$ in the HOMFLY limit $t=q$.
Also, the numerators are the same as in the MacDonald dimensions, but the denominators
are different: this is the obvious origin of the $\gamma$-factors
found in \cite{DMMSS}.

The answer for the generic symmetric representation looks like \cite{FGS1}
\be\boxed{
P_{[r]}^{T[2,2k+1]}(A,q,t)
=\sum_{\ell=0}^r (-1)^{r+\ell}
\frac{(t)_l(q)_r(A)_{r+\ell}(A/t)_{r-\ell}}
{(q)_l(A)_r(qt)_{r+\ell}(q)_{r-\ell}}
 \frac{\{{q}^{2\ell}t\}}{\{t\}}
\left[ {t^lq^{\ell^2-r^2}\over A^r} \right]^{2k+1}}
\ee
An additional symmetry here that generalizes (\ref{symr}) for the superpolynomial is \cite{DMMSS}
\be
P_{[r]}^{T[2,2k+1]}(A,q,t) = P_{[r]}^{T[2,-2k-1]}(A^{-1},q^{-1},t^{-1})
\ =
P_{[1^r]}^{T[2,2k+1]}(A^{-1},t,q)
\ee
This property can be further generalized for the superpolynomial of any torus knot at any representation:
\be
P_R^{T[m,-n]}(A,q,t) = P_R^{T[m,n]}(A^{-1},q^{-1},t^{-1})
\ =
P_{R'}^{T[m,n]}(A^{-1},t,q)
\ee

\subsection{Twist knots}

Our second example is the braids with 2 anti-parallel strands, which describe the twist knots.

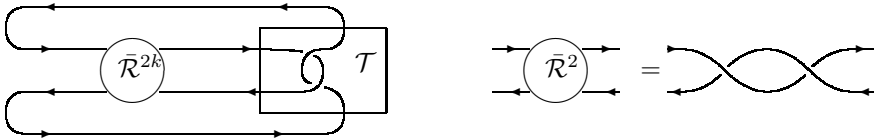
\begin{picture}(200,100)(-100,-50)
\put(0,0){\circle{30}}
\put(-50,-10){\line(1,0){37.5}}
\put(-50,10){\line(1,0){37.5}}
\put(80,-10){\line(-1,0){67.5}}
\put(60,10){\line(-1,0){47.5}}
\put(-50,-30){\line(1,0){140}}
\qbezier(-50,-10)(-60,-10)(-60,-20)
\qbezier(-50,-30)(-60,-30)(-60,-20)
\qbezier(91,-8)(100,-10)(100,-20)
\qbezier(90,-30)(100,-30)(100,-20)
\qbezier(-50,10)(-60,10)(-60,20)
\qbezier(-50,30)(-60,30)(-60,20)
\qbezier(90,10)(100,10)(100,20)
\qbezier(90,30)(100,30)(100,20)
\put(-50,30){\line(1,0){140}}
\qbezier(90,10)(80,10)(80,0)
\qbezier(80,0)(80,-5)(84,-6)
\qbezier(80,-10)(90,-10)(90,0)
\qbezier(90,0)(90,5)(86,7)
\qbezier(80,9)(70,10)(60,10)
\put(-40,30){\vector(-1,0){2}}
\put(-40,10){\vector(1,0){2}}
\put(-40,-10){\vector(-1,0){2}}
\put(-40,-30){\vector(1,0){2}}
\put(70,30){\vector(-1,0){2}}
\put(55,10){\vector(1,0){2}}
\put(55,-10){\vector(-1,0){2}}
\put(70,-30){\vector(1,0){2}}
\put(-7,-3){\mbox{$\bar {\cal R}^{2k}$}}
\put(105,0){\mbox{${\cal T}$}}
\put(60,20){\line(1,0){60}}
\put(60,-20){\line(1,0){60}}
\put(60,-20){\line(0,1){40}}
\put(120,-20){\line(0,1){40}}
\put(200,0){\circle{30}}
\put(170,-10){\line(1,0){17.5}}
\put(170,10){\line(1,0){17.5}}
\put(230,-10){\line(-1,0){17.5}}
\put(230,10){\line(-1,0){17.5}}
\put(195,-3){\mbox{$\bar {\cal R}^2$}}
\put(240,-3){\mbox{$=$}}
\put(253,10){\line(1,0){7}}
\put(253,-10){\line(1,0){7}}
\put(340,10){\line(1,0){10}}
\put(340,-10){\line(1,0){10}}
\qbezier(260,10)(270,10)(280,0)
\qbezier(280,0)(290,-10)(300,-10)
\qbezier(300,10)(310,10)(320,0)
\qbezier(320,0)(330,-10)(340,-10)
\qbezier(260,-10)(270,-10)(278,-2)
\qbezier(281.5,2)(290,10)(300,10)
\qbezier(300,-10)(310,-10)(318,-2)
\qbezier(321.5,2)(330,10)(340,10)
\put(180,10){\vector(1,0){2}}
\put(180,-10){\vector(-1,0){2}}
\put(225,10){\vector(1,0){2}}
\put(225,-10){\vector(-1,0){2}}
\put(256,10){\vector(1,0){2}}
\put(256,-10){\vector(-1,0){2}}
\put(345,10){\vector(1,0){2}}
\put(345,-10){\vector(-1,0){2}}
\end{picture}

\subsubsection{The fundamental representation $R=[1]$\label{tf}}

In this case, in accordance with (\ref{H}), (\ref{evap}), the HOMFLY polynomial is generically given by
\be
T_\fund^{(k)}(A,q) = \alpha_{[1],0}  + \alpha_{[1],1} A^{2k}
\ee
In order to define these two coefficients it is again enough to know the answers
for just {\it two} knots in the series.

For the HOMFLY polynomials of the  {\it ordinary} twist knots such answers
are immediately
available for at least three cases: the unknot $T_\fund^{(0)} = 1$ at any representation, the trefoil (which is a torus
knot, where the Rosso-Jones formula provides an exhaustive answer
at any representation \cite{RJ}) and the figure-eight knot \cite{IMMM},
which can be used to check the answer.  This knowledge implies that
\be\label{atw1}
\alpha_{[1],0} = \frac{A^2}{1-A^2}(q^2+q^{-2} - 1 -A^2) = \frac{A^2}{1-A^2}(1+z^2-A^2),\ \ \ \ \ \ \ \
\alpha_{[1],1}=-\frac{A}{\{A\}}\{Aq\}\{A/q\}
\ee
where $z = q-q^{-1}$. In result,
\be
\boxed{T_\fund^{(k)}(A,q) = 1 + \frac{A^{k+1}\{A^{-k}\}}{\{A\}}\{Aq\}\{A/q\}
\ = 1 -A^{k+1}\{A^k\}\left(\{A\} - \frac{z^2}{\{A\}}\right)= 1 + F_1^{(k)}\{Aq\}\{A/q\}}
\label{felf}
\ee
where we introduced
\be
F_1^{(k)} \equiv \frac{A^{k+1}\{A^{-k}\}}{\{A\}}= A\left(\frac{1}{\{A\}} - \frac{A^{2k}}{\{A\}}\right)
\label{F1}
\ee
Then (comparing with \cite{katlas})
\be
\begin{array}{rll}
&\ldots &\nn
\cr
k=4:& H^{9_2}_\fund(A,q) = T_\fund^{(4)} = 1-\frac{A^5\{A^4\}}{\{A\}}\{Aq\}\{A/q\}
&= (1+z^2)A^2 + z^2A^4 + z^2A^6+ (1+z^2)A^8 - A^{10},\cr
k=3: & H^{7_2}_\fund(A,q) = T_\fund^{(3)} = 1-\frac{A^4\{A^3\}}{\{A\}}\{Aq\}\{A/q\}
&= (1+z^2)A^2 + z^2A^4 + (1+z^2)A^6 - A^8,\cr
k=2: & H^{5_2}_\fund(A,q) = T_\fund^{(2)} = 1-\frac{A^3\{A^2\}}{\{A\}}\{Aq\}\{A/q\}
&= (1+z^2)A^2 + (1+z^2)A^4 -A^6  ,\cr
k=1: & H^{3_1}_\fund(A,q) = T_\fund^{(1)} = 1-A^2\{Aq\}\{A/q\}
&= (2+z^2)A^2 - A^4, \cr
k=0: & H^{\rm unknot}_\fund(A,q) = T_\fund^{(0)}(A)&=1,
\cr
k=-1: & H^{4_1}_\fund(A^{-1},q) = T_\fund^{(-1)}(A) = 1+\{Aq\}\{A/q\}
&= A^{-2} -(1+z^2-A^2) = A^2 - 1-z^2 + A^{-2},
\cr
k=-2: & H^{6_1}_\fund(A^{-1},q) = T_\fund^{(-2)}(A)
= 1+\frac{\{A^2\}}{A\{A\}}\{Aq\}\{A/q\}
&= A^{-4} - A^{-2} + A^2 - z^2(1+A^{-2}),
\cr
k=-3: & H^{8_1}_\fund(A^{-1},q) = T_\fund^{(-3)}(A)
= 1+\frac{\{A^3\}}{A^2\{A\}}\{Aq\}\{A/q\}
& = A^{-6} - (z^2+1)A^{-4}-z^2A^{-2}-z^2+A^2
\cr
k=-4: & H^{10_1}_\fund(A^{-1},q) = T_\fund^{(-4)}(A)
= 1+\frac{\{A^4\}}{A^3\{A\}}\{Aq\}\{A/q\}
&= A^{-8} -(z^2+1)A^{-6}-z^2A^{-4}-z^2A^{-2}-z^2+A^2
\cr
&\ldots&
\end{array}
\ee

The factor $\{A/q\}$ is a must for all symmetric representations,
so that all the answers are equal to unity for $A=q$.
For the antisymmetric representations this is not true.

These HOMFLY polynomials celebrate the recurrent relation:
\be
\boxed{ T_\fund^{(k+1)}(A,q) - A^2T_\fund^{(k)}(A,q) = -A^4 + A^2(1+z^2) }
\ee

\subsubsection{Alternative three-strand ${\cal R}$-matrix calculation\label{3strc}}

The twist knot can be easily represented as a 3-strand knot
with two unusual features, from the point of view of the
general theory of \cite{MMMkn1,MMMkn2}:

1) One of the three strands is in representation $\bar R$

2) Weighted trace is taken only over the states in one strand,
while the combination of two others is instead projected onto the
singlet representation.

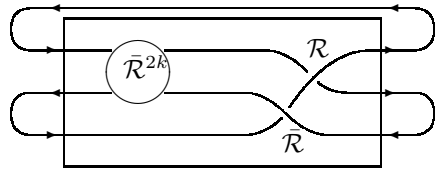
\begin{picture}(200,100)(-180,-50)
\put(0,0){\circle{30}}
\put(-50,-10){\line(1,0){37.5}}
\put(-50,10){\line(1,0){37.5}}
\put(50,-10){\line(-1,0){37.5}}
\put(60,10){\line(-1,0){47.5}}
\put(-50,-30){\line(1,0){100}}
\qbezier(50,-10)(60,-10)(70,-20)
\qbezier(70,-20)(80,-30)(90,-30)
\put(90,-30){\line(1,0){40}}
\qbezier(50,-30)(60,-30)(68,-22)
\qbezier(71.5,-17)(90,10)(110,10)
\put(110,10){\line(1,0){20}}
\qbezier(60,10)(70,10)(80,0)
\qbezier(85,-5)(90,-10)(100,-10)
\put(100,-10){\line(1,0){30}}
\qbezier(-50,-10)(-60,-10)(-60,-20)
\qbezier(-50,-30)(-60,-30)(-60,-20)
\qbezier(130,-10)(140,-10)(140,-20)
\qbezier(130,-30)(140,-30)(140,-20)
\qbezier(-50,10)(-60,10)(-60,20)
\qbezier(-50,30)(-60,30)(-60,20)
\qbezier(130,10)(140,10)(140,20)
\qbezier(130,30)(140,30)(140,20)
\put(-50,30){\line(1,0){180}}
\put(-40,30){\vector(-1,0){2}}
\put(-40,10){\vector(1,0){2}}
\put(-40,-10){\vector(-1,0){2}}
\put(-40,-30){\vector(1,0){2}}
\put(120,30){\vector(-1,0){2}}
\put(120,10){\vector(1,0){2}}
\put(120,-10){\vector(1,0){2}}
\put(120,-30){\vector(-1,0){2}}
\put(-7,-3){\mbox{$\bar {\cal R}^{2k}$}}
\put(68,-38){\mbox{$\bar{\cal R}$}}
\put(80,7){\mbox{${\cal R}$}}
\put(-35,25){\line(1,0){150}}
\put(-35,-45){\line(1,0){150}}
\put(-35,-45){\line(0,1){70}}
\put(115,-45){\line(0,1){70}}
\end{picture}

Since
\be
\l[1]\otimes \overline{[1]} \otimes [1] = \Big([2,1^{N-1}] \oplus [1^N]\Big)\otimes [1]
= [3,1^{N-1}] \oplus 2\times [2,1^N]
\ee
one has two kinds of intermediate representations, and one of them enters
with multiplicity two. Hence, one needs a mixing matrix.
Moreover, there are three different ways to multiply the representations,
\be
\Big(\l[1]\otimes \overline{[1]}\Big) \otimes [1], \ \ \ \
\l[1]\otimes \Big(\overline{[1]} \otimes [1]\Big) \ \ \ \
{\rm and} \ \ \ \
\Big([1]\otimes [1]\Big)\otimes \overline{[1]}
\ee
therefore, there are actually two independent mixing matrices.

These two mixing matrices
in the $[1]\otimes \overline{[1]}$ and $[1]\otimes[1]$ channels are respectively
\be
U_1\sim\left(
\begin{array}{cc}
1 & \sqrt{\{A/q\}\{Aq\}}\\
\sqrt{\{A/q\}\{Aq\}} & -1
\end{array}
\right)
\ee

\be
U_2\sim\left(
\begin{array}{cc}
\sqrt{\{A/q\}} & \sqrt{\{Aq\}}\\
\sqrt{\{Aq\}} & - \sqrt{\{A/q\}}
\end{array}
\right)
\ee

The braid group element is described as (tilde means transposition of the matrix)
\be\label{bg}
B=U_1 {\cal R}_1^n \tilde U_1 {\cal R}_1 U_2 {\cal R}_2^{-1} \tilde U_2
\ee
where the ${\cal R}$-matrices are made of the fundamental representation
eigenvalues in the anti-parallel and parallel 2-strand channels respectively, i.e.
\be
{\cal R}_1\sim\left(
\begin{array}{cc}
1 & 0\\
0 & A
\end{array}
\right)
\ee
\be
{\cal R}_2\sim\left(
\begin{array}{cc}
q/A & 0\\
0 & -1/(qA)
\end{array}
\right)
\ee
Now one has to take the projection onto the singlet state which means to take the element $B_{11}$ in (\ref{bg}).
This element is proportional to $T_\fund^{(k)}(q,A)$.

\subsubsection{The first symmetric representation $R=[2]$}

In this case, the answer is obtained by adjusting the three coefficients
in the expansion over three irreducible representations
with the known ${\cal R}$-matrix eigenvalues (\ref{evap})
with $l=0,1,2$:
\be
T_{[2]}^{(k)}
= \alpha_{[2],0} + \alpha_{[2],1} A^{2k} + \alpha_{[2],2} q^{4k}A^{4k}
= 1 + \frac{\{A/q\}}{\{Aq\}\{A\}}\left\{
\frac{1}{q^2}\{Aq^2\}\Big(A^4q^6 - A^2(q^4-1)(q^2-1)-q^2\Big)- \right.\nn \\
-{A^3\over q^3}(q^2+1)\{Aq\}A^{2k-1}(A^2q^6-q^6+q^4-1)
+ \{Aq^3\}\{A\}^2A^{4k+2}q^{4k+1}\Big\}
\ee
Adjustment is made by comparison with the well-known cases of
the twist knots:
the trefoil $3_1$ ($k=1$), unknot ($k=0$) and the figure-eight knot
$4_1$ ($k=-1$).

This answer can be rewritten in the "canonical" form:
\be
\boxed{T_{_{[2]}}^{(k)} = 1 + F_1^{(k)}
\Big(\{Aq^3\}\{A/q\}  + \{Aq\}\{A/q\}\Big)
+ F_2^{(k)}(Aq)\{Aq^3\}\{Aq^2\}\{A\}\{A/q\}  }
\label{self}
\ee
where the $k$-dependent coefficient $F_1^{(k)}$ "at the first level" remains
the {\bf same} as it was for the fundamental representation!
"At the second level" (i.e. in the third term in (\ref{self})), the $k$-dependence is controlled
by the new peculiar factor
\be
F_2^{(k)}=(A^2q)^{k+1}f_k(Aq) =
A^2q\left(\frac{1}{\{Aq\}\{A\}} - [2]_q\frac{A^{2k}}{\{Aq^2\}\{A\}}
+ \frac{A^{4k}q^{4k}}{\{Aq^2\}\{Aq\}}\right)
=\nn \\
= \frac{qA^2}{\{Aq^2\}\{Aq\}\{A\}}
\Big( \{Aq^2\} - (q+q^{-1})A^{2k}\{Aq\} + q^{4k}A^{4k}\{A\}\Big)
= \nn \\
= \frac{qA^2}{\{Aq^2\}\{Aq\}\{A\}}
\Big( \{Aq^2\} - A^{2k} \{Aq^2\}- A^{2k}\{A\}  + q^{4k}A^{4k}\{A\}\Big)
\label{Ffn2}
\ee
or
\be
f_k(x) = \frac{x^k \{x/q\}\{(xq)^k\} - x^{-k}\{(x/q)^k\}\{xq\}}{\{x/q\}\{x\}\{xq\}}
\label{Fkdef}
\ee
which has the following properties:

\begin{itemize}
\item for all $k$ it is a polynomial in $x^{\pm 1}$ and $q^{\pm 1}$:
at $x=\pm 1$ and $x=\pm q^{\pm 1}$,
when denominator is zero, the numerator also vanishes, and
\be
\ldots \nn \\
f_4(x) = q^{-3}(x^6+x^{-6}) + q^{-2}(q+q^{-1})(x^4+x^{-4})
+ q^{-1}(q^2+1+q^{-2})(x^2+x^{-2}) + (q^3+q+q^{-1}+q^{-3}), \nn \\
f_3(x) = q^{-2}(x^4+x^{-4}) + q^{-1}(q+q^{-1})(x^2+x^{-2}) + (q^2+1+q^{-2}),\nn \\
f_2(x) =q^{-1}(x^2+x^{-2}) + (q+q^{-1}), \nn \\
f_1(x) = 1, \ \ \ f_0(x) = 0, \ \ \ f_{-1}(x) = 1, \nn \\
f_{-2}(x) = q(x^2+x^{-2}) + (q+q^{-1}),\nn \\
f_{-3}(x) = q^{2}(x^4+x^{-4}) + q(q+q^{-1})(x^2+x^{-2}) + (q^2+1+q^{-2}),\nn \\
f_{-4}(x) = q^{3}(x^6+x^{-6}) + q^{2}(q+q^{-1})(x^4+x^{-4})
+ q(q^2+1+q^{-2})(x^2+x^{-2}) + (q^3+q+q^{-1}+q^{-3}), \nn \\
\ldots
\ee
\item all the coefficients of this polynomial are positive integers,
despite it may not look so obvious in (\ref{Ffn2})
\item for $q=1$ this quantity drastically simplifies
and turns into $f_k(x) = \left(\frac{\{x^k\}}{\{x\}}\right)^2$
\item the denominator is a product of three quantities;
actually a "symmetry" between them is also present in the numerator:
it can be rewritten in other forms, containing the products
$\{(xq)^{(k)}\}\{x^{(k)}\}$ or
$\{x^{(k)}\}\{(x/q)^{(k)}\}$ instead of
$\{(xq)^{(k)}\}\{(x/q)^{(k)}\}$,
however, there is no way to make all the three factors present at once:
expression is typically "anomalous".
\end{itemize}

\subsubsection{The next symmetric representation $R=[3]$}

This time $[3]\otimes \overline{[3]}$ contains four
irreducible representations and there will be four
adjustment parameters in the evolution formula:
\be
\boxed{
T_{[3]}^{(k)}
=  \alpha_{[3],0} + \alpha_{[3],1}\cdot A^{2k} + \alpha_{[3],2}\cdot q^{4k}A^{4k} + \alpha_{[3],3}\cdot q^{12k}A^{6k} }
\ee
By itself, the knowledge of trefoil, unknot and figure-eight
is not sufficient anymore: this time one also needs the (known \cite{IMMM3}) HOMFLY polynomial for knot $5_2$.
Presenting the answer in the "canonical" form, one expects now only one new $k$-dependent
structure, at the third level:
\be
T_{[3]}^{(k)}
= 1 + F_1^{(k)}(A,q)
\Big(\{Aq^5\}\{A/q\} + \{Aq^3\}\{A/q\}  + \{Aq\}\{A/q\}\Big) + \nn \\
+ F_2^{(k)}(A,q)\Big(\{Aq^3\}\underline{\{Aq^2\}\{A\}}\{A/q\}+
 \{Aq^5\}\underline{\{Aq^2\}\{A\}}\{A/q\}
+ \{Aq^5\}\underline{\{Aq^4\}\{A\}}\{A/q\}\Big) - \nn \\
+F_3^{(k)}(A,q)\{Aq^5\}\underline{\{Aq^4\}\{Aq^3\}\{Aq\}\{A\}}\{A/q\}
\label{telf}
\ee
where
\be
F_3^{(k)}(A,q) = \frac{A^3q^3}{\{Aq^4\}\{Aq^3\}\{Aq^2\}\{Aq\}\{A\}}\Big(
\{Aq^4\}\{Aq^3\} - (q^2+1+q^{-2})\{Aq^4\}\{Aq\}A^{2k} +\nn\\
+(q^2+1+q^{-2})\{Aq^3\}\{A\}A^{4k}q^{4k} - \{Aq\}\{A\}A^{6k}q^{12k}\Big)
=\nn \\
= A^3q^3\left(\frac{1}{\{Aq^2\}\{Aq\}\{A\}} - [3]_q\frac{A^{2k}}{\{Aq^3\}\{Aq^2\}\{A\}}
+ [3]_q\frac{A^{4k}q^{4k}}{\{Aq^4\}\{Aq^2\}\{Aq\}} -
\frac{A^{6k}q^{12k}}{\{Aq^4\}\{Aq^3\}\{Aq^2\}}\right)
\ee
is a polynomial in $A^{\pm 1}$ and $q^{\pm 1}$.

The terms with $q$-numbers can be rewritten in a variety of ways, for example,
\be
(q^2+1+q^{-2})\{Aq^4\}\{Aq\} =
\{Aq^4\}\Big(\{Aq^3\}+\{Aq\} + \{A/q\}\Big), \nn \\
(q^2+1+q^{-2})\{Aq^3\}\{A\} = \{Aq^4\}\{A/q\} + \{Aq^3\}\{A\} + \{Aq\}\{A\}
\ee

\subsubsection{Generic symmetric representation $R=[r]$ and generic function $F_s^{(k)}$}

The HOMFLY polynomial for arbitrary twist knot $T^{(2k+1)}$ in arbitrary
symmetric representation $[r]$ is given by
\be
\boxed{
T^{(k)}_{[r]} =
1 +  \sum_{s=1}^r \frac{[r]!}{[s]![r-s]!}F_s^{(k)}(A,q)
\prod_{i=1}^s \{Aq^{r+i-1}\}\{Aq^{i-2}\}
} = \nn \\
= 1 + \sum_{s=1}^r F_s^{(k)}(A,q)\cdot
\underline{ \left(\sum_{1\leq j_1<j_2<\ldots
<j_s\leq r}\ \prod_{i=1}^s  \{Aq^{r+j_i-1}\}\right)}\cdot
\prod_{i=1}^{s}\{Aq^{i-2}\}
\label{twigensymm}
\ee
The two formulas are identical because the underlined sum is actually equal to
\be
\sum_{1\leq j_1<j_2<\ldots
<j_s\leq r}\ \prod_{i=1}^s  \{Aq^{r+j_i-1}\}
= \frac{[r]!}{[s]![r-s]!}\prod_{i=1}^s  \{Aq^{r+i-1}\}
\ee
It remains to describe the coefficient function:
\be\label{F}
F_s^{(k)}\!(A^2) = q^{s(s-1)/2}A^s \sum_{j=0}^s
(-)^j\frac{[s]!}{[j]![s-j]!}\frac{ \{Aq^{2j-1}\}\cdot(Aq^{j-1})^{2jk}}{\prod_{i=j-1}^{s+j-1}\{Aq^i\}}
\ee
This completes the description of all symmetric HOMFLY polynomials
for the twist knots.
One can check that these polynomials, indeed, coincide with
those of \cite{indtwist}.

\subsubsection{Superpolynomial in symmetric representation $R=[r]$}

Using formulas (\ref{sevap}) for the eigenvalues and known expressions for the unknot, the trefoil and the figure-eight knot,
one can obtain that in the fundamental case the superpolynomial evolution series
for the twist knots is:
\be
P_\fund^{(k)} = 1+F_1^{(k)}\Big(A^2{q\over t}\Big) \{Aq\}\{A/t\}
\label{pelf}
\ee
Its main property is that
the function $F_1^{(k)}$ does not change, only
its argument changes  (we use $A^2$ as an argument to avoid writing square roots in the change of variables).
This property persists in higher symmetric representations: the functions $F_s^{(k)}(A,q)$, (\ref{F}) remain the same,
just become the functions of $A^2{q\over t}$ and $q$. At the same time, the multipliers $\{Aq^{r+i-1}\}$
in (\ref{twigensymm}) remain the same in the superpolynomial, while $\{Aq^{i-2}\}$ is substituted with
$\{Aq^{i-1}/t\}$. In other words, the answer for the superpolynomial in the generic symmetric representation is

\be
\boxed{
P^{(k)}_{[r]} =
1 +  \sum_{s=1}^r \frac{[r]!}{[s]![r-s]!}F_s^{(k)}(A^2q/t)
\prod_{i=1}^s \{Aq^{r+i-1}\}\{Aq^{i-1}/t\}
} = \nn \\
= 1 + \sum_{s=1}^r F_s^{(k)}(A^2q/t)\cdot
\left(\sum_{1\leq j_1<j_2<\ldots
<j_s\leq r}\ \prod_{i=1}^s  \{Aq^{r+j_i-1}\}\right)\cdot
\prod_{i=1}^{s}\{Aq^{i-1}/t\}
\label{spt}
\ee
Note that the $q$-quantum numbers in this formula are
independent of $t$.
Parameter $t$ enters only in the last factors $\{Aq^{i-1}/t\}$
and through the argument $A^2q/t$ (the binomial coefficients in $F_s^{(k)}$
are again made from the $q$-quantum numbers).
The situation is the opposite for the antisymmetric representations $[1^r]$,
which are described by the mirror change of variables $q \leftrightarrow -t^{-1}$.

\subsection{Two-strand antiparallel links}

After evolution is understood in the channel $R\times \bar R$,
it is easy to evaluate the HOMFLY polynomials for the corresponding
2-strand 2-component braids (see Fig.\ref{2str} with even intersections),
which describe the links of the same topology
as 2-strand torus links, but with inverted direction in one
of the components in Fig.\ref{2str}.

In the case of these antiparallel links,
the eigenvalues are the same as for the twist knots, i.e. in the fundamental representation $R=[1]=\Box$
these are $1$ and
$A$. One may match the general answer
\be
H_{\fund\!\times\bar\fund}^{[2,2k]}(A,q)=\alpha_{\fund,0}+\alpha_{\fund,1} A^{2k}
\ee
with the HOMFLY polynomial of the two untied unknots ($k=0$) and the Hopf link ($k=1$) with antiparallel components. The latter one is
related with the Hopf link with parallel components by the change $A\to 1/A$. In this subsection we consider unreduced
HOMFLY polynomials, which is quite natural just for links, i.e. we let the unknot in the fundamental representation
to be $\{A\}/\{q\}$ (note that the unknot in the representation $\bar\fund$ is equal to the same quantity).
There is no way to fix the normalization of ${\cal R}$-matrix in different components of the link, we choose it so that the
HOMFLY polynomial is equal to unity both at $A=q$ and $A=1/q$.
With these conditions one obtains
\be
H_{\fund\!\times\bar\fund}^{[2,2k]}(A,q)  = 1 + \frac{A^{2k}}{\{q\}^2}\{Aq\}\{A/q\}
\label{invlinksfund}
\ee
in full accordance with \cite{CM} and references therein.

Similarly, in the first symmetric representation, when the unknot both in the representations $[1]$ and $\bar{[1]}$
is equal to ${\{A\}\{Aq\}\over\{q\}\{q^2\}}$,
\be
H_{[2]\times\bar{[2]}}^{[2,2k]}(A,q)  = 1 + \frac{A^{2k}}{\{q\}^2}\{Aq\}\{A/q\} +
\frac{q^{4k}A^{4k}}{\{q\}^2\{q\}^2}\{Aq^3\}\{A\}^2\{A/q\}
\label{invlinksrep2}
\ee

These results imply that the answer in the generic symmetric representation is given just by
\be\label{1}
H_{[r]\times\bar{[r]}}^{[2,2k]}(A,q)=\sum_{i=1}^r\lambda_i^{2k}\check\alpha_i
\ee
Then, the coefficients $\check\alpha_i$ are determined just from the requirement that at $k=0$ (\ref{1}) is equal to the square of
the unknot in representation $r$, i.e. to $\left(\prod_{i=1}^r{\{Aq^{i-1}\}\over\{q^i\}}\right)^2$. Then, the final result is
\be
\check\alpha_i=\left(\prod_{l=1}^{i-1}{\{Aq^{l-1}\}\over\{q^{l+1}\}}\right)^2{\{Aq^{2i-1}\}\{A/q\}\over\{q\}^2}
\ee
i.e.
\be\label{gl}
H_{[r]\times\bar{[r]}}^{[2,2k]}(A,q)=\sum_{i=1}^r\left(\prod_{l=1}^{i-1}{\{Aq^{l-1}\}
\over\{q^{l+1}\}}\right)^2{\{Aq^{2i-1}\}\{A/q\}\over\{q\}^2}A^{ik}q^{ki(i-1)}
\ee
and one can check that the symmetry $k\to -k$, $A\to 1/A$ for properly normalized (\ref{gl}) is satisfied.

\subsection{Double braid knots}

Now we are ready to consider an obvious two-parametric generalization of
torus and twist knots and links, that is, the series of knots which contain the two evolution braids, one with parallel
and another one with antiparallel strands (see ss.\ref{doubleb}).
The series is parameterized by a pair of integers $(m,n)$, and the knots look like

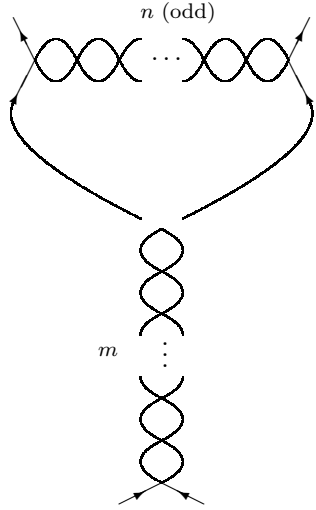
\begin{picture}(200,270)(-230,-240)
\qbezier(-40,0)(-50,20)(-60,0)
\qbezier(-40,0)(-50,-20)(-60,0)
\qbezier(-20,0)(-30,20)(-40,0)
\qbezier(-20,0)(-30,-20)(-40,0)
\qbezier(-20,0)(-15,10)(-10,10)
\qbezier(-20,0)(-15,-10)(-10,-10)
\put(-5,0){\mbox{$\ldots$}}
\qbezier(10,10)(15,10)(20,0)
\qbezier(10,-10)(15,-10)(20,0)
\qbezier(20,0)(30,20)(40,0)
\qbezier(20,0)(30,-20)(40,0)
\qbezier(40,0)(50,20)(60,0)
\qbezier(40,0)(50,-20)(60,0)
\put(-60,0){\line(-1,2){10}}
\put(-60,0){\line(-1,-2){10}}
\put(60,0){\line(1,2){10}}
\put(60,0){\line(1,-2){10}}
\qbezier(0,-80)(-20,-90)(0,-100)
\qbezier(0,-80)(20,-90)(0,-100)
\qbezier(0,-100)(-20,-110)(0,-120)
\qbezier(0,-100)(20,-110)(0,-120)
\qbezier(0,-120)(-10,-125)(-10,-130)
\qbezier(0,-120)(10,-125)(10,-130)
\put(0,-145){\mbox{$\vdots$}}
\qbezier(0,-160)(-10,-155)(-10,-150)
\qbezier(0,-160)(10,-155)(10,-150)
\qbezier(0,-160)(-20,-170)(0,-180)
\qbezier(0,-160)(20,-170)(0,-180)
\qbezier(0,-180)(-20,-190)(0,-200)
\qbezier(0,-180)(20,-190)(0,-200)
\put(0,-80){\line(-2,1){10}}
\put(0,-80){\line(2,1){10}}
\put(0,-200){\line(-2,-1){10}}
\put(0,-200){\line(2,-1){10}}
\put(0,-200){\line(-2,-1){20}}
\put(0,-200){\line(2,-1){20}}
\qbezier(-10,-75)(-80,-40)(-70,-20)
\qbezier(10,-75)(80,-40)(70,-20)
\put(-10,-205){\vector(2,1){2}}
\put(10,-205){\vector(-2,1){2}}
\put(-65,10){\vector(-1,2){2}}
\put(65,10){\vector(1,2){2}}
\put(-70,-20){\vector(1,2){2}}
\put(70,-20){\vector(-1,2){2}}
\put(-10,20){\mbox{\footnotesize$n \ ({\rm odd})$}}
\put(-30,-140){\mbox{\footnotesize $m$}}
%
\end{picture}

$(m,n)$ is a pair such that n is odd (otherwise, this is a link) and m is even (the other orientation is not
consistent). The two-strand torus knots arise for $n=\pm 1$ and the twist knots arise for $m=\pm 2$.
The simplest other knots are: (4,3) is $7_3$, (4,5) is $9_4$, (6,3) is $9_3$ etc. Further:
(4,7): $11a_{342}$ with the braid (1, 1, 1, 1, 1, 2, -1, 2, 3, -2, 3, 4, -3, 4);
(6,5): $11a_{358}$ (1, 1, 1, 1, 1, 1, 1, 2, -1, 2, 3, -2, 3)
(8,3): $11a_{364}$ (1, 1, 1, 1, 1, 1, 1, 1, 1, 2, -1, 2)

In fact, (m,2k+1) describes (k+2)-strand knots with the braid
$(\underbrace{1,...,1}_{m+1},\underbrace{2,-1,2,}_1 \underbrace{3,-2,3,}_2\ldots\underbrace{k+1,-k,k+1}_k)$.

Below we calculate the HOMFLY polynomials for these knots $D^{(m,n)}_r(A,q)$ at any symmetric representation $R=[r]$.
Note that in the course of calculations we use the identities
\be\label{ids}
D^{(0,n)}_r(A,q)=1\nn\\
D^{(2k,\pm 1)}_r(A,q)=X^{(k+1/2\mp 1/2)}_r(A,q)\nn\\
D^{(-2,2k-1)}_r(A,q)=T^{(k)}_r(A,q)\nn\\
D^{(m,n)}_r(A,q)=D^{(-m,-n)}_r(1/A,1/q)
\ee

\subsubsection{The fundamental representation $R=[1]$}

In accordance with (\ref{evp}), (\ref{evap}), the eigenvalues in the fundamental representation are $1/qA$, $q/A$ for the
parallel strands and $1$ and $A$ for the antiparallel strands. This means that the double braid HOMFLY polynomial is of
the form
\be
D^{(m,n)}_{[1]}(A,q)={\alpha_{00}^{[1]}\over (qA)^m}+{\alpha_{01}^{[1]}\cdot q^m\over A^m}+{\alpha_{10}^{[1]}\cdot A^n\over (qA)^m}+
{\alpha_{11}^{[1]}\cdot A^nq^m\over A^m}
\ee
Using (\ref{ids}) and known expression for the torus (\ref{torusf}) knots, one can obtain
\be
\alpha_{00}^{[1]}={\{A/q\}\{q\}\over\{A\}\{q^2\}}\ \ \ \ \ \ \ \ \ \ \ \ \ \ \ \ \ \
\alpha_{01}^{[1]}={\{Aq\}\{q\}\over\{A\}\{q^2\}}\nn\\
\alpha_{10}^{[1]}=-{\{Aq\}\{A/q\}\over\{A\}\{q^2\}}\ \ \ \ \ \ \ \ \ \ \ \ \ \ \ \ \ \
\alpha_{11}^{[1]}={\{Aq\}\{A/q\}\over\{A\}\{q^2\}}
\ee
Note that we obtained the whole series which includes also all the twist knots using the information on only torus knots. This is
since the twist knot sequence is fixed in the fundamental representation by two of twist knots (see ss.\ref{tf}), e.g.
by the unknot and the trefoil which are simultaneously torus and twist knots.

This answer is also reproduced by the 3-strand calculation of ss.\ref{3strc}, one suffices to replace ${\cal R}_2$
with ${\cal R}_2^m$ in (\ref{bg}).

Note that when $A=q^2$, the obtained formulas coincide with (\ref{Jonesmtontw})
directly derived with the help of the Kauffman ${\cal R}$-matrix in the previous section:
\be
q^{2n}D^{(m,n)}(A=q^2)
= \frac{q^{-m}}{D}
\left\{ D\Big(1+q^{2n} - (-q^2)^{-m}q^{2n}\Big)
-  \frac{1+q^{2n}}{D}\Big(1-(-q^2)^{-m}\Big)\right\}
\ee
since $D=q+q^{-1}=\{q^2\}/\{q\}$ and $D^2-1 = q^2+1+q^{-2} = [3]_q\{q\}$.
Note also that (\ref{Jonesmtontw}) provides the unreduced Jones polynomial,
which should be divided by $S_{_\Box}^*(A=q^2) = D = q+q^{-1}$
to be compared with the reduced one $D^{(m;n)}$.
Also, $n$ and $2k+1$ in (\ref{Jonesmtontw})
now are $m$, and $n$ respectively.
Finally, the overall normalization is different
by a factor of $q^{2n} = A^{n}$.

\subsubsection{First symmetric representations}

In the first symmetric representation $R=[2]$
\be
D^{(m,n)}_{[2]}(A,q)=\sum_{ij}^3\alpha_{ij}^{[2]}\left(\lambda_i\right)^n\left(\mu_j^{[2]}\right)^m
\ee
with the eigenvalues given by formulas (\ref{evap}) ($\lambda_i$) and (\ref{evp}) at $r=2$ ($\mu_i$).
Then, using (\ref{ids}) and known expressions for the torus (\ref{torusf}) and twist (\ref{felf}) knots, one obtains
\be
\alpha_{11}^{[2]}={\{A/q\}\{q\}\over \{Aq\}\{q^3\}}\ \ \ \ \ \ \ \ \ \ \ \ \ \ \ \ \ \
\alpha_{12}^{[2]}={\{A/q\}\{Aq^2\}\{q^2\}\over \{Aq\}\{A\}\{q^4\}}\ \ \ \ \ \ \ \ \ \ \ \ \ \ \ \ \ \
\alpha_{13}^{[2]}={\{Aq^2\}\{Aq^3\}\{q\}\{q^2\}\over \{Aq\}\{A\}\{q^3\}\{q^4\}}\nn\\
\alpha_{21}^{[2]}=-{\{A/q\}\over \{q^3\}}\ \ \ \ \ \ \ \ \ \ \ \ \ \ \ \ \ \
\alpha_{22}^{[2]}=-{\{A/q\}\{A^2q^2\}\{q^2\}\over \{Aq\}\{A\}\{q^4\}}\ \ \ \ \ \ \ \ \ \ \ \ \ \ \ \ \ \
\alpha_{23}^{[2]}={\{A/q\}\{Aq^3\}\{q^2\}^2\over \{A\}\{q\}\{q^3\}\{q^4\}}\nn\\
\alpha_{31}^{[2]}={\{A/q\}\{A\}\{Aq^3\}\over \{Aq\}\{q^2\}\{q^3\}}\ \ \ \ \ \ \ \ \ \ \ \ \ \ \ \ \ \
\alpha_{32}^{[2]}=-{\{A/q\}\{A\}\{Aq^3\}\over \{Aq\}\{q\}\{q^4\}}\ \ \ \ \ \ \ \ \ \ \ \ \ \ \ \ \ \
\alpha_{33}^{[2]}={\{A/q\}\{A\}\{Aq^3\}\over \{Aq\}\{q^3\}\{q^4\}}
\ee

Similarly for the next symmetric representation $R=[3]$
\be
D^{(m,n)}_{[3]}(A,q)=\sum_{ij}^4\alpha_{ij}^{[3]}\left(\lambda_i\right)^n\left(\mu_j^{[3]}\right)^m
\ee
with the eigenvalues given by formulas (\ref{evap}) ($\lambda_i$) and (\ref{evp}) at $r=2$ ($\mu_i$). Now, however,
one does not suffices to know the HOMFLY polynomials of the torus and twist knots and use (\ref{ids}). One needs to
know also the HOMFLY polynomials of the three-strand knots (the double braid knots with $n=3$)
in representation $R=[3]$. It can be obtained
from the results of \cite{IMMM3}, and one gets
\be
\alpha_{11}^{[3]}={\{A/q\}\{q\}\over \{Aq^2\}\{q^4\}}\ \ \ \ \ \ \ \ \ \ \ \ \ \ \ \ \ \
\alpha_{12}^{[3]}={\{A/q\}\{Aq^3\}\{q^3\}^2\over \{Aq\}\{Aq^2\}\{q^4\}\{q^5\}}\ \ \ \ \ \ \ \ \ \ \ \ \ \ \ \ \ \
\alpha_{13}^{[3]}={\{A/q\}\{Aq^3\}\{Aq^4\}\{q^2\}\{q^3\}\over \{A\}\{Aq\}\{Aq^2\}\{q^4\}\{q^6\}}\nn\\
\alpha_{14}^{[3]}={\{Aq^3\}\{Aq^4\}\{Aq^5\}\{q\}\{q^2\}\{q^3\}\over \{A\}\{Aq\}\{Aq^2\}\{q^4\}\{q^5\}\{q^6\}}
\ \ \ \ \ \ \
\alpha_{21}^{[3]}=-{\{A/q\}\{Aq\}\over \{Aq^2\}\{q^4\}}\ \ \ \ \ \
\alpha_{22}^{[3]}=-{\{A/q\}\{q^3\}\Big(\{Aq^5\}+\{Aq^3\}-\{A/q\}\Big)\over \{Aq^2\}\{q^4\}\{q^5\}}\nn\\
\alpha_{23}^{[3]}=-{\{A/q\}\{Aq^4\}\{q^2\}\{q^3\}\Big(\{Aq^3\}-\{Aq\}-\{A/q\}\Big)
\over \{A\}\{Aq^2\}\{q\}\{q^4\}\{q^6\}}\ \
\alpha_{24}^{[3]}={\{A/q\}\{Aq^4\}\{Aq^5\}\{q^2\}\{q^3\}^2\over \{A\}\{Aq^2\}\{q\}\{q^4\}\{q^5\}\{q^6\}}\nn\\
\alpha_{31}^{[3]}={\{A/q\}\{A\}\{Aq^3\}\over \{Aq^2\}\{q^2\}\{q^4\}}\ \ \ \ \
\alpha_{32}^{[3]}={\{A\}\{A/q\}\{Aq^3\}\{q^3\}\Big(\{Aq^5\}-\{Aq\}-\{A/q\}\Big)
\over \{Aq\}\{Aq^2\}\{q^2\}\{q^4\}\{q^5\}}\nn\\
\alpha_{33}^{[3]}=-{\{A\}\{A/q\}\{Aq^3\}\{q^3\}\Big(\{Aq^5\}+\{Aq^3\}-\{Aq\}\Big)
\over \{Aq\}\{Aq^2\}\{q\}\{q^4\}\{q^6\}}\ \ \ \ \ \  \
\alpha_{34}^{[3]}={\{A/q\}\{A\}\{Aq^3\}\{Aq^5\}\{q^3\}^2\over \{Aq\}\{Aq^2\}\{q\}\{q^4\}\{q^5\}\{q^6\}}\nn\\
\alpha_{41}^{[3]}=-{\{A/q\}\{A\}\{Aq\}\{Aq^5\}\over \{Aq^2\}\{q^2\}\{q^3\}\{q^4\}}\ \ \ \ \ \ \ \ \ \ \ \ \ \ \ \ \ \
\alpha_{42}^{[3]}={\{A/q\}\{A\}\{Aq\}\{Aq^5\}\{q^3\}\over \{Aq^2\}\{q\}\{q^2\}\{q^4\}\{q^5\}}\nn\\
\alpha_{43}^{[3]}=-{\{A/q\}\{A\}\{Aq\}\{Aq^5\}\over \{Aq^2\}\{q\}\{q^4\}\{q^6\}}\ \ \ \ \ \ \ \ \ \ \ \ \ \ \ \ \ \
\alpha_{44}^{[3]}={\{A/q\}\{A\}\{Aq\}\{Aq^5\}\over \{Aq^2\}\{q^4\}\{q^5\}\{q^6\}}
\ee

\subsubsection{Generic symmetric representation $R=[r]$}

Similarly, one can construct the HOMFLY polynomials for higher symmetric representation in order to anticipate an answer
in the generic symmetric representation $R=[r]$. Thus, we look for the HOMFLY polynomial of the double braid knot in the form
\be\boxed{
D^{(m,n)}_r(A,q)=\sum_{s,i,j=0}^rF^{(r)}_{si}X^{(r)}_{sj}\Big(\lambda_i\Big)^n\Big(\mu^{(r)}_j\Big)^m
=\sum_{s,i,j=0}^rF^{(r)}_{si}X^{(r)}_{sj}A^{in-rm}q^{mi(i-1)+nj(j+1)-mr^2}
}\label{dbg}
\ee
with the eigenvalues given by formulas (\ref{evap}) ($\lambda_i$) and (\ref{evp}) ($\mu_i$).
The answer is better to formulate in a form manifestly splitting
the dependencies on the torus and twist braids. To this end, we represent the HOMFLY polynomials of the
twist knots (\ref{twigensymm}) in the form
\be
T^{(k)}_r(A,q)=\sum_{s,l=0}^r F^{(r)}_{sl}\lambda_l^{2l+1}
\ee
with $\lambda_i$ from (\ref{evap}) and the sum over $s$ is the same as in (\ref{twigensymm}), i.e.
\fr{
F^{(r)}_{kl}=(-)^l{\Big(A/q\Big)_k\Big(q^r\Big)^*_k\Big(Aq^r\Big)_k\over
\Big(Aq^{l-1}\Big)_{k+1}\Big(q\Big)_l\Big(q\Big)_{k-l}}A^{l+k}q^{k(k-1)/2+l(l-1)}\{Aq^{2l-1}\}\ \ \ \ \ \ \hbox{at}\ k\ge l\nn\\
F^{(r)}_{kl}=0\ \ \ \ \ \ \hbox{at}\ k<l
}
Similarly, the toric braid
contribution is described by the coefficients $X^{(r)}_{sj}$. We look for these coefficients in the form
\be\boxed{
X^{(r)}_{sj}={(-)^{s}\over A^{r+s}\{q\}^r}P_{r-s,r-j}\ \ \ \ \ \ \ \ \ \
P_{sj}=\sum_{i=0}^s\alpha^{(r)}_{sij}\Big(Aq^{2r-j-1}\Big)_{s-i}^*\Big(Aq^{j-2}\Big)^*_i}
\ee
where $\alpha^{(r)}_{sij}$ are some coefficients which we determine by the procedure described above. They turn out to be
\fr{
\alpha^{(r)}_{sij}={1\over
q^{i^2+s^2/2-si+sj-2ij+(2r+1)i-(4r-1)s/2+3r(r-1)/2}}\cdot{1\over
\{q\}^{j-s-1}}\times\nn\\
\times{\Big(q^s\Big)^*_i\Big(q^{r-s}\Big)^*_{j-i}\over
\Big(q^{2r-2j+i+1}\Big)^*_{s+1}}{[r-j+i]![2r-2j+1][2r-2j-s+2i+1]
\over [i]![j-i]![2r-s-j+i+1]!}
}
Note that from the first identity in (\ref{ids}) and (\ref{dbg}) follows that
\be
\sum_{j=0}^rX_{sj}^{(r)}=\delta_{s0}
\ee

\section{The differential hierarchy of colored knot polynomials}

Now, possessing a systematic description of relatively
large families, we can try to use them for revealing additional
structures, which are {\it a priori} not so obvious.
Of these we concentrate on the hierarchy of "differentials",
which seem to be a rather general feature of at least HOMFLY
colored polynomials, in, at least, symmetric representations.
It generalizes the structure first found in \cite{IMMM}
in the figure-eight knot $4_1$
(see \cite{AnoMMM21} for its further
generalization to non-symmetric representations):
we demonstrate that it remains almost the same for
all twist knots
(actually, this was already noted in \cite{indtwist})
and, in just a slightly weaker form,
for a considerably larger variety of knots
(what was probably anticipated in \cite{DGR}).
This structure is not just found by analyzing the results
obtained by the evolution method, it seems to be somehow
deeply consistent with the evolution method. However, what this exactly means
is yet somewhat difficult to formulate.

\bigskip

To begin with, in the Abelian case $A=q\ $ the HOMFLY polynomial $H_{_\Box}(A,q)$ reduces to
a single monomial $q^{2s_{+}}$.
Due to the "mirror" symmetry \cite{DMMSS,GS}
\be
(R,A,t,q) \rightarrow (\bar R,A,-q^{-1},-t^{-1}),
\label{mirsym}
\ee
where $\bar R$ is the transposed Young diagram $R$,
the same is true for $H_{_\Box}(A,q^{-1})$,
i.e. at $A=q^{-1}$
the original $H_{_\Box}(A,q)$ is also a monomial
$q^{2s_{-}}$.
In fact, the symmetry implies more: $s_+=s_-=0$,\ i.e.
\be
H_{_\Box}(A,q) = 1 + G_1(A,q)\{Aq\}\{A/q\}
\label{H1diff}
\ee
The examples can be found in this paper: (\ref{felft}), (\ref{felf}), (\ref{invlinksfund}).
This is often obscured by the wrong normalization of the HOMFLY polynomial,
but with the properly introduced "normalization factor" (that is, corresponding to the
topological framing, see ss.3.1)
this seems to be always true.
Clearly, if $s_+$ and $s_-$ are non-vanishing, one can multiply the HOMFLY
polynomial with a factor, and $q^{-s_--s_+} A^{s_--s_+} H_{\Box}$
would have the property (\ref{H1diff}):
in fact this is exactly the factor that
provides the correct topological-invariant normalization.

For the colored ($R\neq\Box$) knot polynomials above, the factorization of $H-1$
becomes a little more involved.
For the group $SU_q(N)$
(associated with the point $A=q^N$ or, better, $A=t^N$)
the representation $R$ with $l(R)>N$
lines is trivial, while for $l(R)=N$  it is equivalent to
a representation with $l(R)=N-1$ lines:
the rule is $R=\{r_1\geq r_2\geq \ldots \geq r_N\geq 0\}
\cong R'=\{r_1-r_N\geq r_2-r_N\geq\ldots r_{N-1}-r_N\geq 0\}$.
For instance, the pure antisymmetric representation
$[1^N] = \overline{[N]}$ is equivalent to a singlet,
i.e. behaves as if it was the Abelian case,
in particular, $H_{[1^N]}$ is a monomial of $q$ when $A=q^r$.
This implies that in the dual symmetric representation
$H_{[r]}$ should be a monomial when $A=q^{-r}$.
One could conclude in the same way that for $A=q^{-r''}$
with $r''<r$ it simply vanishes. HOWEVER, this applies to
THE {\it unreduced} polynomial $\check H_{[r]}$: it indeed vanishes,
but so does the quantum dimension $D_{r''}$ and nothing
special happens to THE {\it reduced} POLYNOMIAL $H_{[r]}$.
For the same reason $H_R$ turns into A monomial only for
$A=q^{l(R)}$, but for the symmetric representation $[r]$ this means that
the special point is exactly $A=q$.
Finally, with the properly chosen framing,
the both monomials are just unities, and, as a generalization of
(\ref{H1diff}), one has for the HOMFLY polynomials
in symmetric and antisymmetric representations
\be
\begin{array}{ccc}
H_{[r]} = 1 + \tilde G_r(A,q)\{Aq^r\}\{A/q\} \\ & & H_{[r]}(A=q) = 1  \\
\updownarrow \ {\rm mirror}  & \Longleftarrow \\
 & & H_{[1^r]}(A=q^r)=1\\
H_{[1^r]} = 1 + \tilde G_r(A,q^{-1})\{Aq\}\{A/q^r\}
\end{array}
\label{Hrdiff}
\ee
Since the two formulas are of course related by the symmetry
(\ref{mirsym}),   in what follows we consider only the
first one.
For the generic Young diagrams $R$ the above reasoning implies
somewhat weaker statements:
\be
H_R - H_{R'} \sim \{A/q^{l(R)}\}, \nn \\
H_{\bar R} - H_{\bar R'} \sim \{A/q^{l(\bar R)}\}
\label{HRdiffw}
\ee
see \cite{AnoMMM21} for more details and applications
of these properties.

Thus, one see that the property (\ref{Hrdiff}) is
a simple and natural consequence of "general principles",
hence, it can be easily generalized to the superpolynomials:
if the $t$-deformation respects the general properties
of $SU_q(N)$ representation theory
(what is widely
believed to be true, despite the relevant deformation
of the group structure is far more involved than just
an ordinary quantum group),
then one expects that
\be
P_{[r]} = 1 + \tilde{\cal G}_r(A,q,t) \{Aq^r\}\{A/t\}, \nn \\
P_{[1^r]} = 1 + (-1)^{r+1}\tilde{\cal G}_r(A,-t^{-1},-q^{-1})\{Aq\}\{A/t^r\}
\ee
(again, see \cite{AnoMMM21} for a discussion of generic
representations $R$).

\bigskip

What is far more interesting, $\tilde G_r$ in (\ref{Hrdiff})
has a non-trivial {\bf additional} structure.
If one subtracts from $\tilde G_r$ the first $G_1$ from
(\ref{H1diff}) with an appropriate knot-independent
coefficient, then the difference factorizes further:
\be
\tilde G_r - [r]_q\tilde G_1 = \tilde{\tilde G}_r\{Aq^{r+1}\}
\label{Hrdiff2}
\ee
We already saw this phenomenon for the figure-eight knot
in \cite{IMMM}.
Today we know that it holds at least for all the 3-strand knots
($m=3$, not only $m=2$\ !) from
\cite{IMMM2,IMMM3}
where higher symmetric representations are available,
and a natural conjecture is that it is true for {\it all} knots
Moreover, (\ref{Hrdiff2}) continues to a whole hierarchy of
embedded factorizations, just like in the figure-eight case:
\be
\begin{array}{ccc}
H_{_\Box} = H_{[1]} =  & 1 + G_1(A,q)\{Aq\}\{A/q\},& \cr
H_{[2]} = &1 + (q+q^{-1}) G_1(A,q)\{Aq^2\}\{A/q\}
&+\, G_2(A,q)\{Aq^3\}\{Aq^2\}\{A/q\}, \cr
H_{[3]} = &1 + (q^2+1+q^{-2}) G_1(A,q)\{Aq^3\}\{A/q\}
&+ (q^2+1+q^{-2}) G_2(A,q)\{Aq^4\}\{Aq^3\}\{A/q\} + \cr
&&\ \ \ \ \
+ G_3(A,q)\{Aq^5\}\{Aq^4\}\{Aq^3\}\{A/q\}, \cr
\ldots&&
\end{array}
\ee
and in general
\be
\boxed{
H_{[r]}^{\cal K} = 1 + \sum_{j=1}^r \frac{[r]!}{[j]![r-j]!}\,
G_j^{\cal K}(A,q) \left(\prod_{i=0}^{j-1} \{Aq^{r+i}\}\right)\{A/q\}
}
\label{Hrdiffh}
\ee
In fact, according to \cite{IMMM} for the figure-eight knot the
factorization (\ref{Hrdiff2}) is even deeper:
the r.h.s. contains a factor
$\{Aq^{r+1}\}\{Aq/t\} \ \stackrel{t=q}{=}\ \{Aq^{r+1}\}\{A\} = Z_{r|1}^{(1)}$.
As we demonstrated in this paper, this property persists for
all twist knots (including the trefoil), moreover, entire (\ref{Hrdiffh})
for the twist knots is enhanced to a $Z$-expansion:
\be
H_{[r]}^{twist} = 1 + \sum_{j=1}^r \frac{[r]!}{[j]![r-j]!}\,
F_j^{twist}(A^2,q^2) \left(\prod_{i=0}^{j-1} \{Aq^{r+i}\}\{Aq^{i-1}\}\right)
\label{Htwrdiffh}
\ee
where the factors come in pairs, $Z_{r|1}^{(i)} = \{Aq^{r+i}\}\{Aq^{i-1}\}$.
However, beyond this family
only the weaker factorizations (\ref{Hrdiff2}) and (\ref{Hrdiffh}) take place,
still they do!

\bigskip

Since the combinations $\{Aq^a/t^b\}$ are nothing but the DGR differentials
introduced in \cite{DGR}, we call (\ref{Hrdiffh}) "differential hierarchy".
In particular, in \cite{IMMM} this (actually, enhanced) structure
was used to derive an equation relating different symmetric
representation: a counterpart of the $A$-polynomial, but a different one.

\bigskip

The structure (\ref{Hrdiffh}) should be respected by the
$t$-deformation to the superpolynomials, and, indeed, it is, as we shall
also see in the examples below.
For instance, for the twist knots
\be
\boxed{
P_{[r]}^{twist} = 1 + \sum_{j=1}^r \frac{[r]!}{[j]![r-j]!}\,
F_j^{twist}(A^2q/t,\,q^2) \left(\prod_{i=0}^{j-1} \{Aq^{r+i}\}\{Aq^i/t\}\right)
}
\label{Ptwrdiffh}
\ee
with just the same functions $F_j$ as in (\ref{Htwrdiffh}).
This formula coincides with the result of \cite{indtwist},
but is structured in a different way.
Of course, the $t$-deformation rule can not be and is not
just so simple for multi-strand knots, still the structure
(\ref{Hrdiffh}) seems very useful for that purposes.
At least the two rules are definitely universal:
\begin{itemize}
\item the differentials change in a regular way:
$\{Aq^{r+i}\}$ remain intact, while  $\{Aq^i/q\} \longrightarrow \{Aq^i/t\}$,
in particular, $Z^{(i)}_{r|1} \longrightarrow {\cal Z}^{(i|0)}_{r|1}
=\{Aq^{r+i}\}\{Aq^i/t\}$,
\item in coefficient functions like $F$ and $G$ all $A^2\longrightarrow A^2q/t$.

\noindent
What remains a question is:
\item how $q^2$ is changed in the coefficient functions; at the moment
this seems to depend on the set of knots/links, e.g. for the twisted knots
$q^2$ remains intact, while for the 2-strand torus knots rather $q^2\longrightarrow qt$.
\end{itemize}

\section{Conclusion. The new boost for the $Z$-expansion}

In this paper we presented a broad review of the evolution method,
concentrating on the case of the 2-strand ($m=2$) braids with
parallel and anti-parallel orientation of strands. These later
include the 2-strand torus knots ($3_1,5_1,7_1,\ldots$), the twist
knots ($4_1, 5_2, 6_1, 7_2, \ldots$), counter-oriented 2-strand
links and some other families, for example the ones, containing the
knots $7_3$, $9_3$, $9_4$, $\ldots$
or $8_3$, $10_3$, $\ldots$.
Not only we reproduce the
known answers for the (anti)symmetric representations of the twist \cite{IMMM,indtwist} and
some torus \cite{FGS1} knots and links, but add a new important example of a
2-parametric family of double braids.
Many more examples with $m\geq 3$ could
be added, with smaller $m$ providing initial conditions for larger
$m$, however, we do not yet possess a clear formulation of this hierarchy.

In general, the evolution method allows one to obtain entire one- or
multi-parametric sets of answers for varieties of knots and
representations, and this helps to reveal certain hidden
structures, which are the traces of an underlying integrable
structure involving the colored HOMFLY polynomials of all knots and
links, and which still has to be discovered. In this section, we briefly
summarize what is achieved on this route in the present paper.

The most profound is the structure found for the twisted knots: it
turns out that what was first discovered in \cite{IMMM} for the
figure-eight knot $4_1$ remains true for the entire family; to see this one
suffices to look at the results of \cite{indtwist}
from the proper angle. Namely, for arbitrary, say, symmetric
representation $[r]$
\be
T^{(k)}_{[r]} = 1 + \sum_{s=1}^r
F^{(k)}_s\Big(A^2,q^2\Big)\prod_{i=1}^{s} Z_{r|1}^{(i-1)}
\label{Texp}
\ee
Following \cite{IMMM},
we call this representation of knot polynomials "$Z$-expansion".
What is important here,

(i) the coefficient functions $F^{(k)}_s(A^2,q^2)$ are independent
of $r$, i.e. characterize the knot itself,

(ii) as a further manifestation of this, the answer for the colored ($t$-deformed)
superpolynomial immediately follows:
\be
{\cal T}^{(k)}_{[r]} = 1 + \sum_{s=1}^r
F^{(k)}_s\!\!\left(\frac{A^2q}{t},q^2\right)\prod_{i=1}^{s} {\cal Z}_{r|1}^{(i-1)}
\ee
Within the set of the twisted knots the two obvious next questions are

(iii) generalizations to arbitrary representations: here the approach of
\cite{AnoMMM21,AnoMor,AnoMMM} seems promising but needs much
more work; and

(iv) better understanding of the $s$-dependence of
the coefficient functions: since (\ref{Texp})
is actually a $q$-deformation of the archetypical
relation \cite{DMMSS}
\be
\sigma^{\cal K}_R(A) = \Big(\sigma_{_\Box}^{\cal K}(A)\Big)^{|R|}
\ee
for the special polynomials, all $F_s$'s should be somehow constructed from
the $q$-binomial coefficients and $F_1$,
this, with further various advances on $Z$-expansion, is discussed
in the forthcoming paper \cite{Art}.

\bigskip

In this paper we moved in another direction: to other
families of knots and links.
For the 2-strand torus knots, it is
impossible to preserve (\ref{Texp}) with all its nice properties. If
one insists on (i), $r$-independent coefficients arise only in
front of much "weaker" products, $Z_{r,1}\prod_{i=1}^{s-1}
\{Aq^{r+i}\}$, see (\ref{Htwrdiffh}).
If one insist on $Z$-expansion as it is, then the
coefficient functions depend on $r$, and the shifts $i$ are not
restricted to $s$; this option is discussed in great detail in \cite{Art}.

To further emphasize the role of the coefficients functions, we
derived another archetypical formula (\ref{dbg}), demonstrating that
{\bf when one makes a kind of a {\it combination} of the two braids,
the corresponding coefficient functions are a kind of convoluted}:
this seems to be a deep and far-going generalization of the
multiplication property of knot polynomials for the {\it composite}
knots (see, e.g., \cite[eq.(44)]{MMMkn2}).

Thus, not only this paper reviews a powerful and important
{\it evolution method}, as a byproduct result we hope to give a new
boost to a very different approach: that of the $Z$-expansion of
\cite{IMMM}. In \cite{Art} even more is claimed:
that the coefficient functions of the $Z$-expansion can be considered
as a prototype for a clever coordinates in the space of all knots and links.
This seems to be quite an interesting suggestion,
but it could hardly be supported without a variety of examples,
provided, in particular, by the {\it evolution method}, though this
does not at all exhaust its potential, and at least for a while
will remain an important technical and conceptual tool in the study of
knot polynomials.

\section*{Acknowledgements}

We are grateful to S.Artamonov, I.Cherednik, E.Gorsky, D.Melnikov and S.Shakirov
for valuable discussions.

Our work is partly done within the framework of the SFFR of Ukraine Grant No.
F53.2/028, partly supported by Ministry of Education and Science of
the Russian Federation under contract 8410 and by the Dynasty Foundation (And.Mor.), by the Brazil National Counsel
of Scientific and Technological Development (A.Mor.), by
NSh-3349.2012.2, by RFBR grants 13-02-00457 (A.Mir.), 13-02-00478
(A.Mor.) and 11-02-01220 (And.Mor.), by joint grants 12-02-92108-Yaf-a,
13-02-91371-ST-a.


\begin{thebibliography}{12}

\bibitem{knotpol} J.W.Alexander, 
Trans.Amer.Math.Soc. {\bf 30} (2) (1928) 275?306;\\
J.H.Conway, 
Algebraic Properties, In: John Leech (ed.), {\sl Computational
Problems in Abstract Algebra}, Proc. Conf. Oxford, 1967, Pergamon
Press, Oxford-New York, 329-358, 1970;\\
V.F.R.Jones, 
Invent.Math. {\bf 72} (1983) 1
Bull.AMS {\bf 12} (1985) 103
Ann.Math. {\bf 126} (1987) 335;\\
L.Kauffman,
Topology {\bf 26} (1987) 395;\\
P.Freyd, D.Yetter, J.Hoste, W.B.R.Lickorish, K.Millet,
A.Ocneanu,
Bull. AMS. {\bf 12} (1985) 239;\\
J.H.Przytycki and K.P.Traczyk, 
Kobe J. Math. {\bf 4} (1987) 115-139

\bibitem{CS} S.-S.Chern and J.Simons,
Ann.Math. {\bf 99} (1974) 48-69

\bibitem{WitJones} A.S.Schwarz, New topological invariants arising in the theory of quantized fields,
Baku Topol. Conf., 1987;\\
E.Witten, Comm.Math.Phys. {\bf 121} (1989) 351

\bibitem{TR} E.Guadagnini, M.Martellini and M.Mintchev, Clausthal 1989,
Procs.
 307-317;
Phys.Lett. {\bf B235} (1990) 275;\\
N.Yu.Reshetikhin and V.G.Turaev, 
Comm. Math. Phys. {\bf 127} (1990) 1-26

\bibitem{MMMkn1} A.Mironov, A.Morozov and And.Morozov, arXiv:1112.5754

\bibitem{MMMkn2} A.Mironov, A.Morozov and And.Morozov, JHEP {\bf 03} (2012)
034, arXiv:1112.2654

\bibitem{AS} M.Aganagic and Sh.Shakirov, arXiv:1105.5117; arXiv:1210.2733

\bibitem{DMMSS} P.Dunin-Barkowski, A.Mironov, A.Morozov, A.Sleptsov and
A.Smirnov, JHEP {\bf 03} (2013) 021,
arXiv:1106.4305

\bibitem{RJ} M.Rosso and V.F.R.Jones, J. Knot Theory Ramifications, {\bf 2}
(1993) 97-112

\bibitem{HLsuper} A.Mironov, A.Morozov, Sh.Shakirov and A.Sleptsov, JHEP {\bf 2012} (2012) 70,
arXiv:1201.3339

\bibitem{Che} I.Cherednik, arXiv:1111.6195

\bibitem{net} A.Negut, arXiv:1209.4242;\\
E.Gorsky and A.Negut, arXiv:1304.3328

\bibitem{IMMM} H.Itoyama, A.Mironov, A.Morozov and And.Morozov, JHEP {\bf 2012} (2012) 131, arXiv:1203.5978

\bibitem{indtwist} S.Nawata, P.Ramadevi, Zodinmawia and X.Sun, arXiv:1209.1409

\bibitem{FGSS} H.Fuji, S.Gukov, M.Stosic and P.Sulkowski, arXiv:1209.1416

\bibitem{MMeqs} A.Mironov and A.Morozov,  AIP Conf.Proc. {\bf 1483} (2012) 189-211, arXiv:1208.2282

\bibitem{CJ} I.Goulden and D.Jackson,
Proc.Amer.Math.Soc. \textbf{125} (1997) 51-60,
math/9903094

\bibitem{MMN} A.Mironov, A.Morozov and S.Natanzon,  Theor.Math.Phys. {\bf 166} (2011) 1-22,
arXiv:0904.4227;  Journal of Geometry and Physics {\bf 62} (2012) 148-155,
arXiv:1012.0433

\bibitem{chi} X.-S.Lin and H.Zheng, Trans. Amer. Math. Soc. {\bf 362} (2010) 1-18 math/0601267;\\
S.Stevan,
Annales Henri Poincar\'e {\bf 11} (2010) 1201-1224,
arXiv:1003.2861

\bibitem{BEM}  A.Brini, B.Eynard and M.Mari\~no, arXiv:1105.2012

\bibitem{Kaufm} L.Kauffman, Topology {\bf 26} (1987) 395-407; Trans.Amer.Math.Soc. {\bf 311} (1989) 697-710;\\
L.Kauffman and P.Vogel, J.Knot Theory Ramifications {\bf 1} (1992) 59-104

\bibitem{KoMamo} M. Kontsevich, 
Funkts. Anal. Prilozh., \textbf{25:2} (1991) 50-57;
Comm.Math.Phys. {\bf 147} (1992) 1-23; \\
S.Kharchev, A.Marshakov, A.Mironov, A.Morozov and A.Zabrodin,
Phys. Lett. \textbf{B275} (1992) 311-314, hep-th/9111037;
Nucl.Phys. \textbf{B380} (1992) 181-240, hep-th/9201013; \\
P.Di Francesco, C.Itzykson and J.-B.Zuber,
Comm.Math.Phys. {\bf 151} (1993) 193-219, hep-th/9206090

\bibitem{HLhomfly} A.Mironov, A.Morozov and Sh.Shakirov, J. Phys. A: Math. Theor. {\bf 45} (2012) 355202, arXiv:1203.0667

\bibitem{Sham} Sh.Shakirov et al, {\it to appear}

\bibitem{IMMM2} H.Itoyama, A.Mironov, A.Morozov, And.Morozov,
Int.J.Mod.Phys. {\bf A27} (2012) 1250099,
arXiv:1204.4785

\bibitem{DM} V.Dolotin and A.Morozov, JHEP {\bf 1301} (2013) 065, arXiv:1208.4994;  arXiv:1209.5109

\bibitem{FGS1} H.Fuji, S.Gukov and P.Sulkowski,  arXiv:1203.2182

\bibitem{katlas} Knot Atlas at http://katlas.org/wiki/Main Page (by D.Bar-Natan)

\bibitem{CM} N.Carqueville and D.Murfet, arXiv:1108.1081

\bibitem{IMMM3} H.Itoyama, A.Mironov, A.Morozov, And.Morozov,
Int.J.Mod.Phys. {\bf A28} (2013) 1340009,
arXiv:1209.6304

\bibitem{AnoMMM21} A.Anokhina, A.Mironov, A.Morozov and An.Morozov,  arXiv:1211.6375

\bibitem{GS}  S.Gukov and M.Stosic,  arXiv:1112.0030

\bibitem{DGR} N.M.Dunfield, S.Gukov and J.Rasmussen, Experimental Math. 15
(2006) 129-159,
math/0505662

\bibitem{AnoMor} A.Anokhina and And.Morozov, {\it to appear}

\bibitem{AnoMMM} A.Anokhina, A.Mironov, A.Morozov and And.Morozov, Nucl.Phys. {\bf B868} (2013) 271-313,
arXiv:1207.0279; arXiv:1304.1486

\bibitem{Art} S.Artamonov et al, {\it to appear}

\end{thebibliography}
\end{document}